\documentclass[10pt]{article}
\usepackage{amsmath,amssymb,graphicx,color}

\usepackage{geometry}
\usepackage{natbib}

\makeatletter

\DeclareRobustCommand\citepos
  {\begingroup
   \let\NAT@nmfmt\NAT@posfmt
   \NAT@swafalse\let\NAT@ctype\z@\NAT@partrue
   \@ifstar{\NAT@fulltrue\NAT@citetp}{\NAT@fullfalse\NAT@citetp}}

\let\NAT@orig@nmfmt\NAT@nmfmt
\def\NAT@posfmt#1{\NAT@orig@nmfmt{#1's}}

\makeatother

\newcommand{\be}{\begin{equation}}
\newcommand{\ee}{\end{equation}}

\newcommand{\bfb}{\boldsymbol{b}}

\newcommand{\bfd}{\boldsymbol{d}}
\newcommand{\bfe}{\boldsymbol{e}}

\newcommand{\bfn}{\boldsymbol{n}}
\newcommand{\bfp}{\boldsymbol{p}}

\newcommand{\bfr}{\boldsymbol{r}}
\newcommand{\bft}{\boldsymbol{t}}
\newcommand{\bfu}{\boldsymbol{u}}
\newcommand{\bfv}{\boldsymbol{v}}
\newcommand{\bfx}{\boldsymbol{x}}

\newcommand{\bfI}{\boldsymbol{I}}

\newcommand{\bfP}{\boldsymbol{P}}

\newcommand{\bfxi}{\boldsymbol{\xi}}

\newcommand{\bfka}{\boldsymbol{\kappa}}

\newcommand{\half}{\textstyle{\frac12}}

\newcommand{\grads}{\nabla_{\mskip-2mu\scriptscriptstyle S}}

\newcommand{\divs}{\text{div}_{\mskip-2mu\scriptscriptstyle S}}
\newcommand{\tr}{\text{tr}\mskip2mu}

\newcommand{\sdot}{\mskip-1.25mu\cdot\mskip-1.25mu}
\newcommand{\stimes}{\mskip-1.5mu\times\mskip-1.25mu}

\newcommand{\dr}{\,\text{d}r}
\newcommand{\dtheta}{\,\text{d}\theta}
\newcommand{\drdtheta}{\,\text{d}r\mskip1mu\text{d}\theta}

\newcommand{\Dr}[1]{\frac{\text{d}#1}{\text{d}r}}

\newcommand{\lint}{\int_0^{2\pi}}
\newcommand{\aint}{\int_0^{2\pi}\mskip-8mu\int_0^1}

\newcommand{\xir}{\bfxi_r}

\newcommand{\xit}{\bfxi_{\theta}}
\newcommand{\xitt}{\bfxi_{\theta \theta}}
\newcommand{\xittt}{\bfxi_{\theta \theta \theta}}

\newcommand{\vt}{\bfv_\theta}
\newcommand{\vtt}{\bfv_{\theta\theta}}

\newcommand{\subthet}{\mskip-4mu\theta}

\newcommand{\sperp}{\scriptscriptstyle\perp\mskip-4mu}

\defcitealias{Haijun}{Zhou \& Ou-Yang (1999)}

\renewcommand{\eta}{\nu}

\newcommand{\captionfonts}{\footnotesize}

\makeatletter  
\long\def\@makecaption#1#2{%
  \vskip\abovecaptionskip
  \sbox\@tempboxa{{\captionfonts #1: #2}}%
  \ifdim \wd\@tempboxa >\hsize
    {\captionfonts #1: #2\par}
  \else
    \hbox to\hsize{\hfil\box\@tempboxa\hfil}%
  \fi
  \vskip\belowcaptionskip}
\makeatother   

\begin{document}

\title{Stability and buckling of flat circular configurations of closed, intrinsically nonrectilinear filaments spanned by fluid films}

\author{T.\ M.\ Hoang \& Eliot Fried
\\[12pt]
\small\emph{Mathematical Soft Matter Unit}\\[-2pt]
\small\emph{Okinawa Institute of Science and Technology Graduate University}\\[-2pt]
\small\emph{Onna, Okinawa 904-0415, Japan}}

\date{}

\maketitle

\begin{abstract}
\noindent
A variational model is used to study the behavior of flexible but inextensible loops spanned by liquid films, with the objective of explaining the stability and buckling of flat circular configurations. Loops made from filaments with intrinsic curvature and/or intrinsic twist density are considered, but it is assumed that those loops have circular cross sections and uniform mechanical properties. 
%
%
For a loop made from a filament with intrinsic curvature but no intrinsic twist density, there exist in-plane and out-of-plane buckling modes corresponding to stable solution branches that bifurcate from the branch of flat circular solutions. In this case, out-of-plane buckling occurs at a lower value of the dimensionless surface tension of the fluid film than does in-plane buckling. Additionally, however, the destabilizing influence of the intrinsic curvature can be countered by increasing the torsional rigidity of the filament relative to its flexural rigidity. For a loop made from a filament with both intrinsic curvature and intrinsic twist density, only one branch of stable solutions bifurcates from the flat circular solution branch. In this case, the in-plane and out-of-plane buckling modes are intertwined and bifurcation occurs at a value of the dimensionless surface tension less than that governing the behavior of loops made from filaments that are intrinsically rectilinear. Furthermore, increasing the torsional rigidity of the filament relative to its flexural rigidity has no or little stabilizing effect if the loop is either too short or too long. Besides, in contrast to what occurs for a loop made from a filament with no intrinsic twist density, the destabilizing influence of the intrinsic curvature cannot be countered by increasing the torsional rigidity of the filament relative to its flexural rigidity if the intrinsic twist density of the filament is sufficiently large, regardless of the length of the loop. 
\end{abstract}

\section{Introduction}
\label{introduction}

A surface embedded in three-dimensional point space is said to be minimal if, given any simple closed curve on the surface, the area of the portion of the surface surrounded by that curve is less than that of any 
contiguous surface with the same boundary. 
%
Granted sufficient smoothness, the related problem of finding the surface of least area that spans a given, simple, closed space curve leads to the requirement that the mean curvature of the surface must vanish pointwise. The mathematical challenges associated with that boundary-value problem have been of long-standing interest and have inspired major advances in differential geometry, real and complex analysis, partial-differential equations, and the calculus of variations. 
Moreover, results concerning the existence, uniqueness, and qualitative properties of solutions to the problem have direct applications in architecture \citep{otto,emmer}, image processing \citep{fuchs,johnstone}, computer-aided design \citep{monterde,xu}, and various other fields. Granted sufficient regularity, solutions to the classical problem provide non-trivial geometries for static, non-compact, black brane horizons in asymptotically flat space-time \citep{emparan,armas}.


In his groundbreaking studies of capillarity, \cite{plateau} observed that analog solutions to the previously mentioned boundary-value problem are provided by the liquid films that form after closed wire frames are dipped into and extracted from soapy water. This observation applies to films with lateral dimensions that are orders of magnitude larger than the typical thickness of a wet soap film, which is approximately $10^2$ nm, and is a direct consequence of force balance. To understand the role of force balance, consider two regions at possibly different uniform pressures separated by a surface $S$ oriented by a unit normal vector $\bfn$ and tension $\sigma>0$. Force balance then requires that, at each point of $S$,
\be
\grads\sigma+(2\sigma\mskip-2muH-p)\bfn=\bf0,
\label{yleqn}
\ee
where $\grads$ is the gradient operator on $S$, $H=-\half\divs\bfn$ is the mean curvature of $S$ (with $\divs$ being the divergence operator on $S$), and $p$ is the constant difference between the pressure in the region into which $\bfn$ points and the region out of which $\bfn$ points. The tangential and normal components of \eqref{yleqn}, respectively, imply that, for $S$ to be in equilibrium, $\sigma$ and $H$ must obey
\be
\sigma=\text{constant}
\qquad\text{and}\qquad
H=\frac{p}{2\sigma}=\text{constant}.
\label{yleqn2}
\ee
Although the pressure inside a soap bubble must always exceed that outside the bubble, the pressure cannot differ across an openly suspended soap film. By \eqref{yleqn2}, any surface $S$ that serves as a model for a soap film must have pointwise vanishing mean curvature and therefore be minimal. 


An interesting alternative to the problem of finding a minimal surface that spans a given, simple, closed-space curve arises if the boundary is flexible, while being both inextensible and elastically resistant to bending. That problem was first considered by \cite{bernatzki} and \cite{bernatzkiye}, who established the existence of solutions for boundaries without self-contact. Motivated by experiments with loops of fishing line and soapy water, \cite{giomi} focused on characterizing the shapes that form for different values of the dimensionless surface tension, which scales with the cube of the radius of the loop, and which can therefore be tuned simply by changing the length of the loop without changing its mechanical properties or the composition of the soap solution. \cite{chen} then studied the stability and bifurcation of flat circular solutions, after which \cite{aisa1,aisa2} extended their results to situations where the boundary curve resists not only bending but also twisting.


Since commercially available fishing line is generally spooled for transport and storage, it invariably exhibits intrinsic curvature. Additionally, it is quite difficult to join the ends of a strand of fishing line to create loops that are free of intrinsic twist density. Motivated by these observations, our goal here is to explore how the stability and buckling behavior of a flexible, but inextensible loop spanned by a liquid film in a flat circular configuration, is influenced by the intrinsic shape of the fiber from which the loop is made. 

The papers of \cite{Michell}, \citetalias{Haijun}, and \cite{Goriely1} 
provide important precedents for our work. \cite{Michell} considered the influence of intrinsic twist density on the stability of a twisted elastic ring and derived an expression for the critical twist, 
in terms of the ratio of torsional rigidity to flexural rigidity, above which the ring twists out of its plane. Subsequently, \citetalias{Haijun} extended \citepos{Michell} approach to the linearized dynamics of rings to incorporate the effect of intrinsic curvature. They discovered an instability that arises if the actual radius of curvature of the ring is greater than its intrinsic radius of curvature, and on this basis they provided an explanation for the kinking of DNA minicircles. Using a different approach based on obtaining nontrivial period solutions of the relevant linearized equilibrium equations, \cite{Goriely1} also considered and arrived at conclusions consistent with those of \citetalias{Haijun}. Our work can be viewed as an effort to extend the contributions of \cite{Michell}, \citetalias{Haijun}, and \cite{Goriely1} to account for the force that a spanning liquid film exerts on a filament with intrinsic curvature and intrinsic twist density.

The results of \cite{giomi}, \cite{chen}, and \cite{aisa1,aisa2} show that a circular loop that is made from an intrinsically rectilinear filament with circular cross section and is spanned by a liquid film bifurcates stably to an in-plane configuration at a critical value of the dimensionless surface tension. They also show that a stable bifurcation from a circular configuration to an out-of-plane configuration is not possible. We discover that intrinsic curvature and intrinsic twist density can influence not only the value of the dimensionless surface tension at which the flat circular configuration becomes unstable but also the buckling mode that accompanies the ensuing bifurcation. We also quantify these effects. In a related effort, \cite{Giulio} recently considered the influence of the intrinsic curvature on the stability and buckling behavior of a circular loop of noncircular cross-section, focusing on situations where the radius of the loop is less thank its intrinsic radius of curvature. They found that intrinsic curvature must be present to ensure the existence of a first buckling mode that is not purely tangent to the spanning surface. 
By way of contrast, we consider a circular loop of circular cross-section with both intrinsic curvature and intrinsic twist density but impose no restriction between the intrinsic curvature of the loop and its radius.

The remainder of this paper is organized as follows. Our basic modeling assumptions are described in Section~\ref{preliminaries}.
The first and second variations of the net potential energy that embodies these assumptions are derived in Section~\ref{variationconditions}. A parametric reformulation of the problem is provided in Section~\ref{parametricrecasting}. Stability and buckling analyses appear in Sections~\ref{stability} and Section~\ref{buckling}, respectively. 
A summary of our most important results appears in Section~\ref{summary}, which concludes with a brief discussion of the limitations of our model and potential directions for future work.

\section{Preliminaries}
\label{preliminaries}

We consider a flexible loop spanned by a liquid film. We model the filament from which the loop is made as an inextensible and unshearable rod with midline $C$. In addition to assuming that the rod is endowed with resistance to bending and twisting, we allow for the possibility that its rest configuration possesses intrinsic curvature and/or intrinsic twist density. As we note in the Introduction, intrinsic curvature is a common feature of commercially available filaments and intrinsic twist density is easily induced when joining the free ends of a segment of filament to make a (closed) loop. Additionally, we model the liquid film as a surface $S$ endowed with resistance to local area changes in the form of a uniform surface tension.
%
Additionally, we assume that the system consisting of the loop and the liquid film is isolated, so that its total free-energy coincides with its total potential-energy.

\begin{figure}[t]
\centering
\includegraphics[width=0.975\textwidth]{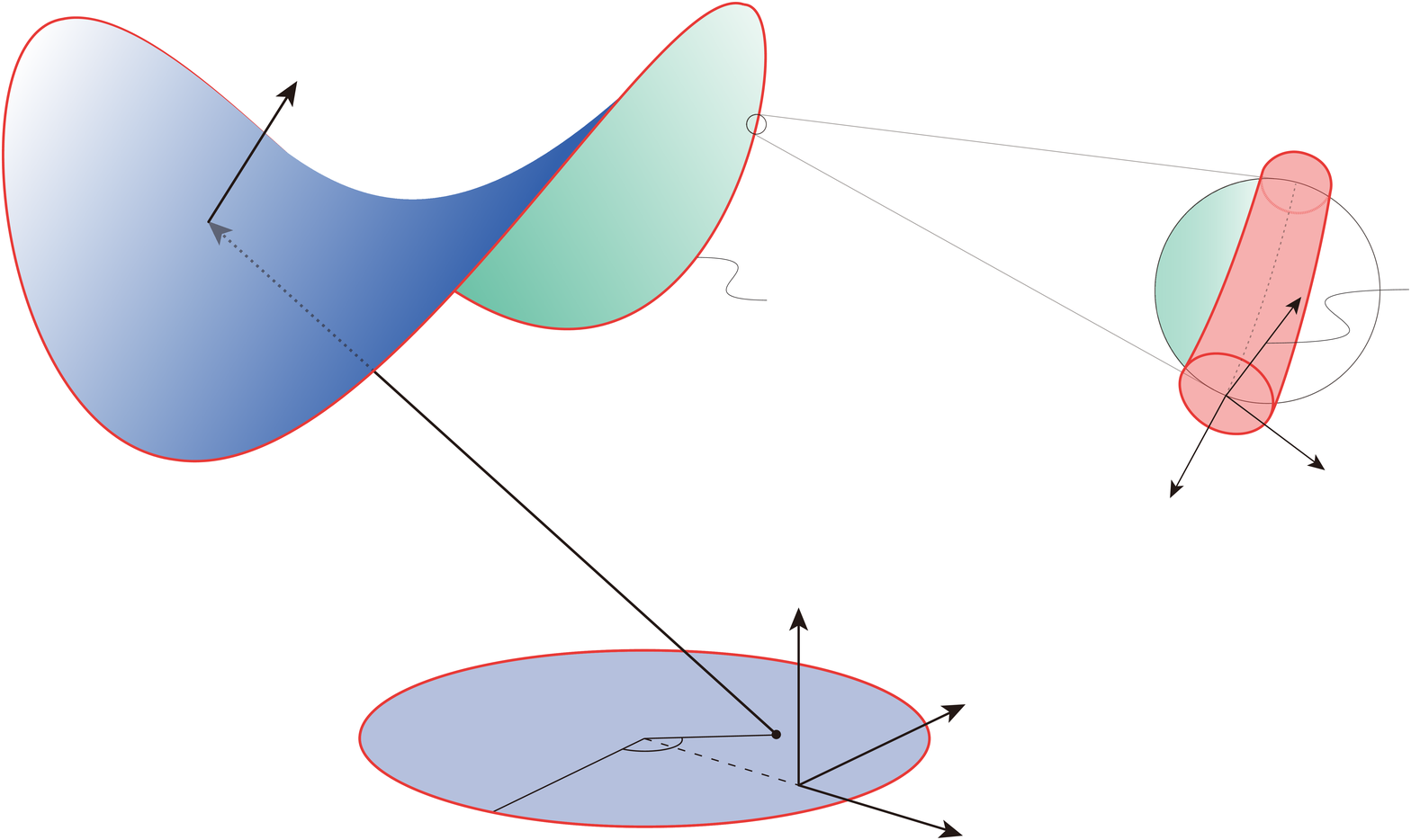}
\put(-273,42){\small$1$}
\put(-230.5,51.25){\small$r$}
\put(-245,37.5){\small$\theta$}
\put(-155,18.25){\small$\bfe_r$}
\put(-154.5,58.5){\small$\bfe_\theta$}
\put(-207,89.5){\small$\bfe_z$}
\put(-400,230){\small$S$}
\put(-210,171){\small$C=\partial S$}
\put(-358,241){\small$\bfn(r,\theta)$}
\put(-101,109){\small$\bft$}
\put(-52,120){\small$\bfd$}
\put(-26.5,174){\small$\bft\times\bfd$}
\put(-276,116){\small$R\bfxi(r,\theta)$}
\caption{A fluid film represented by an orientable surface $S=\{\bfx: \bfx=R\bfxi(r,\theta),0\le r \le 1, 0 \le \theta \le 2\pi\}$. The film spans a flexible loop made from a filament with centerline $C=\{\bfx: \bfx=R\bfxi(1,\theta), 0 \le \theta \le 2\pi\}$ coincident with the boundary $\partial S$ of $S$. The surface $S$ has unit normal $\bfn$ and the space curve $C$ is endowed with a triad $\{\bft,\bfd,\bft \times \bfd\}$ of orthogonal directors, with $\bft$ being tangent to $C$ and $\bfd$ residing in the normal cross-section of the filament.}
\label{model}
\end{figure}
We restrict attention to situations where the system is configured such that the curve $C$ has no points of self-contact and the surface $S$ is orientable. Moreover, we stipulate that the midline $C$ of the rod and the boundary $\partial S$ of $S$ coincide, so that, as depicted schematically in Figure~\ref{model},
\be
C=\partial S.
\label{coincidence}
\ee

\subsection{Kinematics}
\label{kinematics}

We let $\bft$ denote a unit vector field to $C$ and introduce a director field $\bfd$ satisfying
\be
|\bfd|=1
\qquad\text{and}\qquad
\bfd\cdot\bft=0.
\label{dconstraints}
\ee
Consistent with the special Cosserat theory of rods, as presented by \cite{Antman}, the orthonormal triad $\{\bft,\bfd,\bft\times\bfd\}$ provides a material frame for $C$. By \eqref{dconstraints}$_2$, $\bfd$ lies in the cross-section of the rod and admits a representation
\be
\bfd=\cos\psi\mskip2mu\bfp+\sin\psi\mskip2mu\bfb
\label{drepresentation}
\ee
in terms of the normal and binormal elements $\bfp$ and $\bfb$ of the Frenet frame $\{\bft,\bfp,\bfb\}$ of $C$. From \eqref{drepresentation}, $\psi$ gives the angle by which $\bfp$ must be rotated, about $\bft$, to coincide with $\bfd$. 

Using a prime to denote differentiation with respect to arclength along $C$, we define the curvature vector $\bfka$ and the twist density $\omega$ of $C$ by
\be
\bfka=\bft^\prime
\qquad\text{and}\qquad
\omega=(\bfd\times\bfd^\prime)\cdot\bft.
\label{kvecandomega}
\ee
By \eqref{drepresentation} and the Frenet--Serret relations,
\be
\bft^\prime = \kappa \bfp, \qquad \bfp^\prime = -\kappa \bft + \tau \bfb, \qquad \bfb^\prime = -\tau \bfp,
\label{FSrlns}
\ee
where $\tau$ defined according to
\be
\kappa^2 \tau = (\bft \times \bft') \cdot \bft''
\label{taudef}
\ee
is the torsion of $C$, \eqref{kvecandomega}$_2$ yields
\be
\omega = \psi^\prime+\tau.
\label{omegarel}
\ee

We introduce a unit normal field $\bfn$ for $S$. The mean curvature $H$ of $S$ is then given by
\be
H=-\frac{1}{2}\divs\bfn,
\label{meancurvaturedef}
\ee 
where $\divs$ denotes the surface divergence operator on $S$. Following \cite{giomi}, it is useful to introduce the angle $\vartheta$ between the Frenet normal $\bfp$ of $C$ and the restriction to $C$ of $\bfn$. In terms of $\vartheta$, the restriction to $C$ of $\bfn$ then takes the form 
\be
\bfn = \cos\vartheta\mskip2mu\bfp + \sin\vartheta\mskip2mu\bfb.
\ee

\subsection{Energetics}
\label{energetics}

We take the total free-energy of the system to be of the particular form
\be
E=\int_C \frac{1}{2}(a|\bfka-\kappa_0\bfd |^2 + c(\omega-\omega_0)^2)
+\int_S\sigma,
\label{E}
\ee
where $a>0$ and $c>0$ are the constant bending and twisting rigidities of the rod, $\kappa_0\ge0$ and $\omega_0$ are the constant intrinsic curvature and intrinsic twist density, and $\sigma>0$ is the constant surface tension of the soap film. 

For $\kappa_0=0$ and $\omega_0=0$, \eqref{E} reduces to the expression considered by \cite{aisa1,aisa2}. Otherwise, \eqref{E} involves a vectorial measure $\bfka-\kappa_0\bfd$ of bending strain and a scalar measure $\omega-\omega_0$ of twisting strain. Whereas the second of these measures is conventional, the former is less so. In contrast to the scalar measure $\kappa-\kappa_0$ of bending strain, which appears in many models for intrinsically curved filaments,  $\bfka-\kappa_0\bfd$ accounts for coupling between bending of the midline and cross-sectional orientation. In particular, since $\bfka\cdot\bfd=\kappa\cos\psi$ by \eqref{drepresentation}, \eqref{kvecandomega}$_1$, and \eqref{FSrlns}$_1$, the bending energy per unit length associated with $\bfka-\kappa_0\bfd$ is proportional to $(\kappa-\kappa_0)^2 +2\kappa_0 \kappa(1-\cos\psi)$ and consequently penalizes unfavorable combinations of bending and cross-sectional orientation.

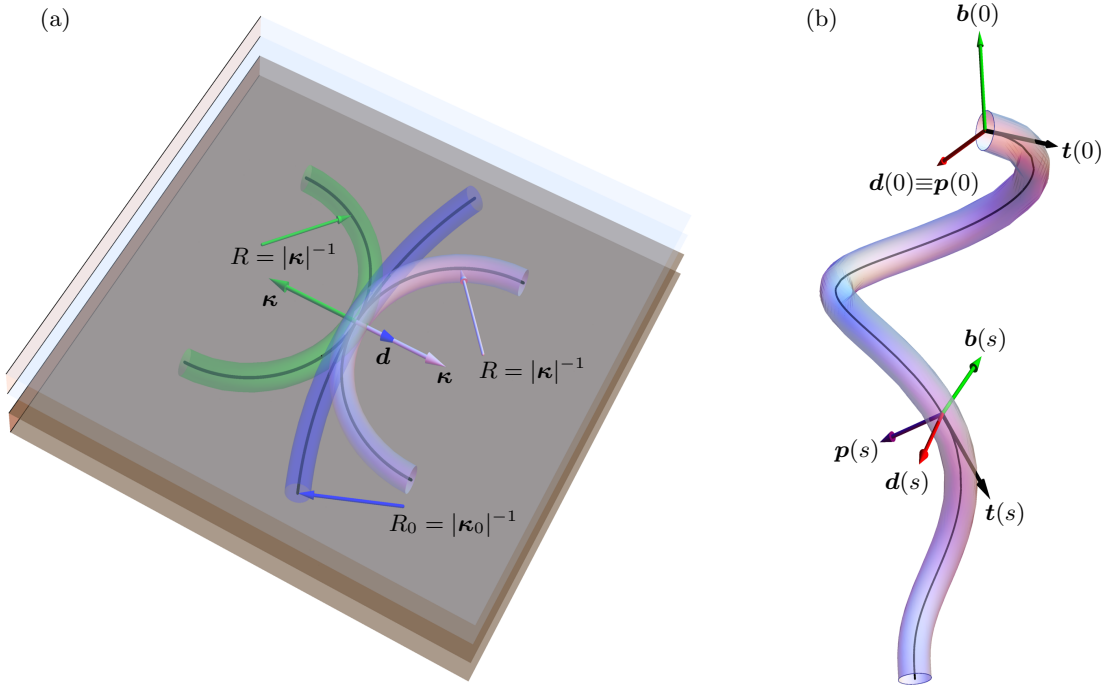
\begin{figure}[t]
\centering
\includegraphics[width = 95mm,height=95mm]{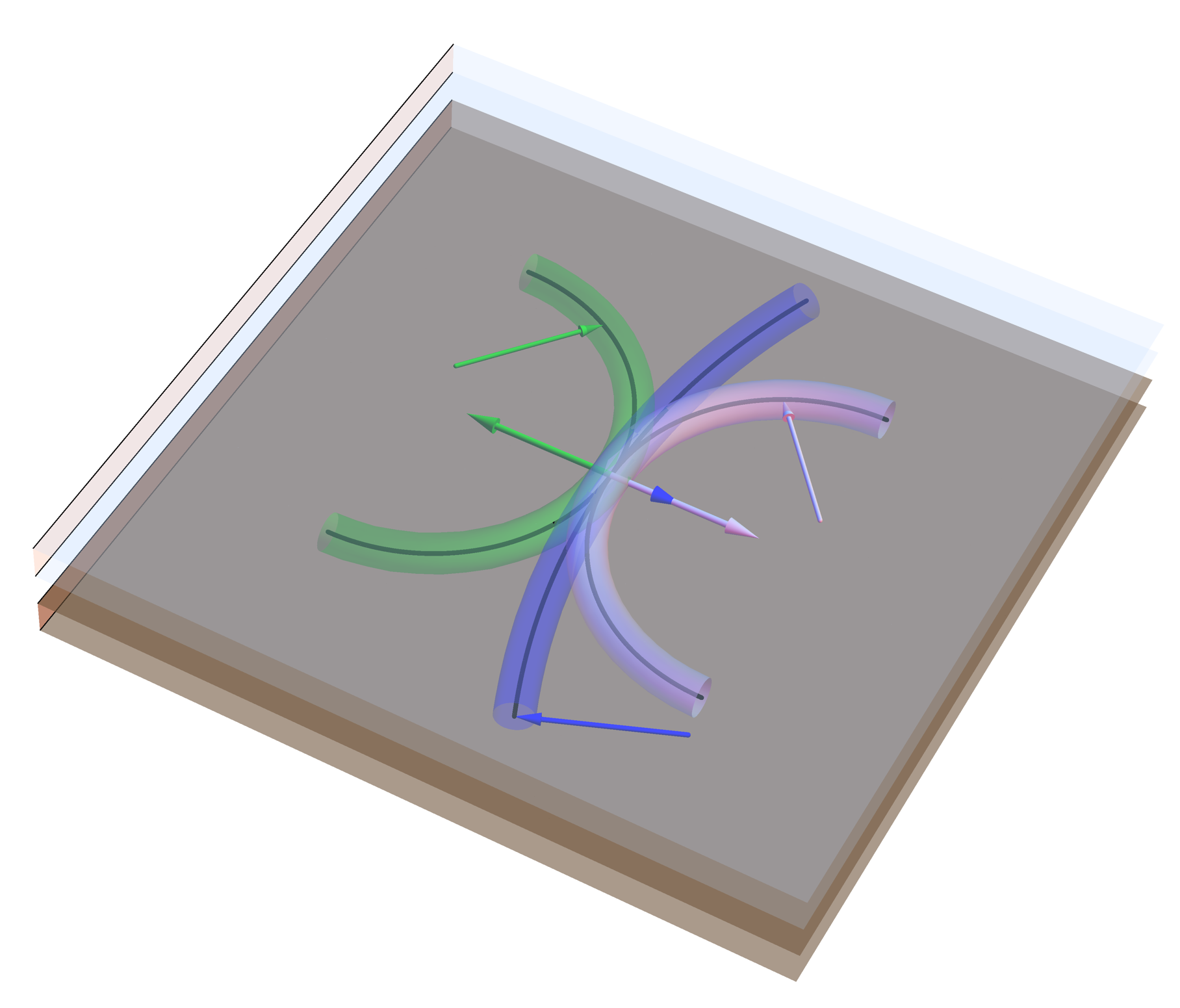}
\put(-84,122){\small $R=|\bfka|^{-1}$}
\put(-178,165){\small $R=|\bfka|^{-1}$}
\put(-118,64){\small $R_0=|\bfka_0|^{-1}$}
\put(-100,120){\small $\bfka$}
\put(-166,149){\small $\bfka$}
\put(-123,127){\small $\bfd$}
\put(-250,255){\small $(\text{a})$}
\qquad\qquad
\includegraphics[width = 35mm,height=90mm]{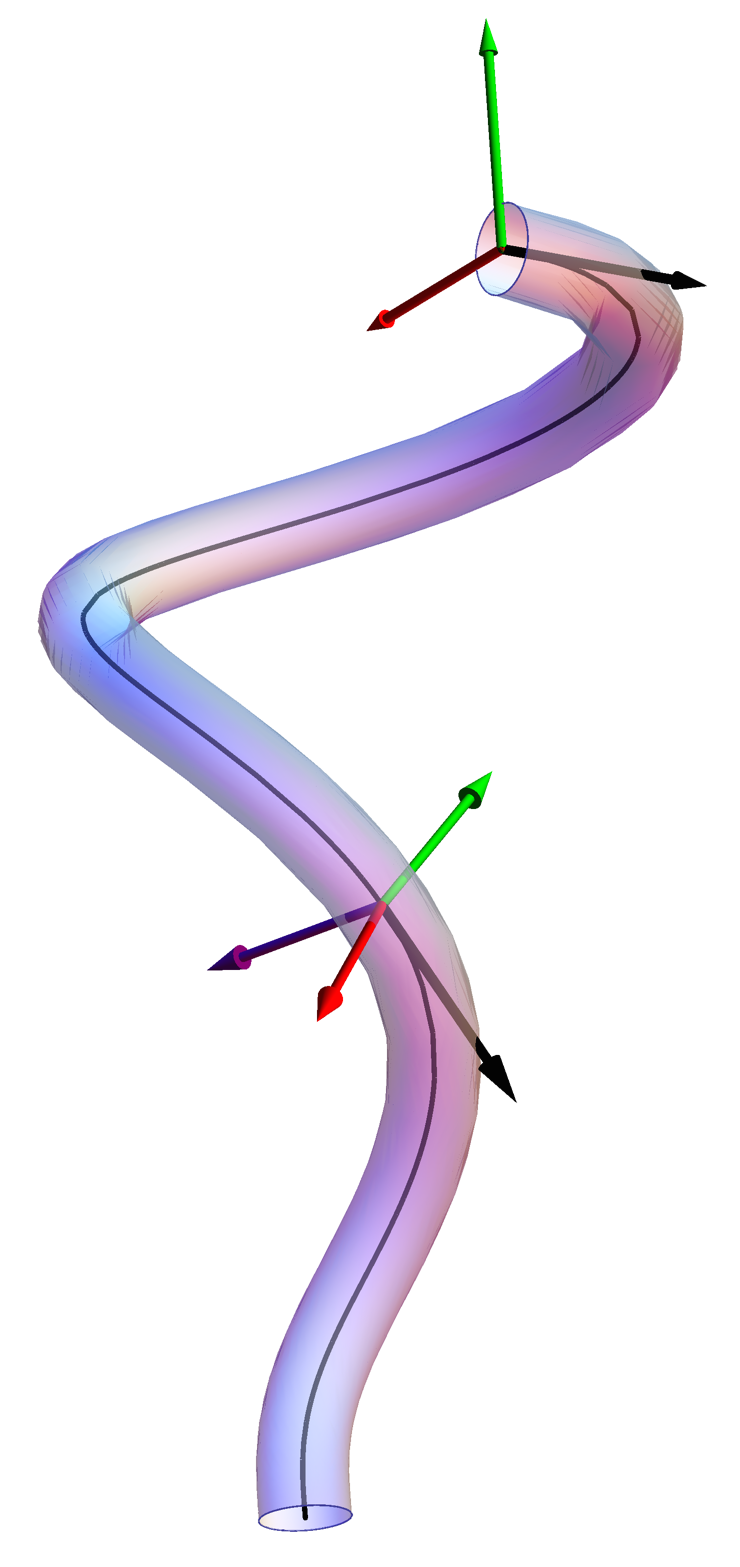}
\put(-43,257){\small$\bfb(0)$}
\put(-76,193){\small$\bfd(0)$$\equiv$$\bfp(0)$}
\put(-3,205){\small$\bft(0)$}
\put(-40,134){\small$\bfb(s)$}
\put(-89,92){\small$\bfp(s)$}
\put(-70,80){\small$\bfd(s)$}
\put(-32,68){\small$\bft(s)$}
\put(-100,255){\small $(\text{b})$}
\caption{Illustration of bending strain of a filament with intrinsic curvature $\kappa_0$ for (a) planar bending without rotating the director $\bfd$ about the midline of the filament, in which case the vector curvature $\bfka$ aligns with $\bfd$ and (b) bending with the magnitude $\kappa=|\bfka|$ of the vector curvature constant and equal to the intrinsic curvature $\kappa_0$.} 
\label{fig2}
\end{figure}
An important feature of this penalization becomes evident in the restricted context of purely planar bending. To see this, consider a filament with intrinsic curvature $\kappa_0$ that is confined between two parallel plates with a gap slightly greater than the diameter of the filament. There are two ways of bending the filament to achieve a circular shape with uniform curvature $\kappa_*\ne\kappa_0$, depending on whether its ends are moved toward or away from the center of intrinsic curvature (Figure~\ref{fig2}a). In only the second of these alternatives is there a tendency for the cross sections of the filament to rotate away from the center of intrinsic curvature toward the center of curvature induced by bending. The scalar measure $\kappa-\kappa_0$ of  bending strain is insensitive to the distinction between these alternatives. As a consequence, the associated bending energy per unit length, which is proportional to $(\kappa-\kappa_0)^2$, does not account for the additional work needed to rotate the cross sections to achieve a configuration bent away from the center of intrinsic curvature. In contrast, $\bfka-\kappa_0\bfd$ differs for in-plane bending toward or away from the center of intrinsic curvature. Specifically, for $\bfka=\kappa_*\cos\psi\mskip2mu\bfd$, the strain measure $\bfka-\kappa_0\bfd=(\kappa_*\cos\psi-\kappa_0)\bfd$ reduces to $(\kappa_*-\kappa_0)\bfd$ for $\psi=0$ and to $-(\kappa_*+\kappa_0)\bfd$ for $\psi=\pi$ and therefore yields bending energy densities respectively proportional to $(\kappa_*-\kappa_0)^2$ and $(\kappa_*+\kappa_0)^2$.

Further insight concerning the nature of the vectorial strain measure $\bfka-\kappa_0\bfd$ arises on considering a nonplanar configuration in which the magnitude of the vector curvature is required to be constant with magnitude $\kappa_0$ while rotating about the tangent of the midline (Figure~\ref{fig2}b). Under these circumstances, the scalar measure $\kappa-\kappa_0$ of bending strain vanishes, as does the associated bending energy per unit length. This seems difficult to justify. In contrast, the bending energy per unit length associated with the vectorial measure $\bfka-\kappa_0\bfd$ of bending strain yields is proportional to $|\bfka-\kappa_0\bfd|^2=2\kappa_0^2(1-\cos\psi)$, which vanishes only if $\psi=0$. The work needed to rotate the cross-section about $\bft$ an angle $\psi$ for $\bfd$ to coincide with $\bfp$ is therefore taken into account. 

Next, using \eqref{drepresentation} and the orthonormality of the Frenet frame to express the normal element $\bfp$ of that frame by $\bfp=\cos\psi\mskip2mu\bfd-\sin\psi\mskip2mu\bft\times\bfd$, noting from \eqref{kvecandomega}$_1$ and \eqref{FSrlns}$_1$ that $\bfka=\kappa\bfp$, and introducing scalar measures of curvature 
\be
\kappa_1=\kappa\cos\psi
\qquad\text{and}\qquad
\kappa_2=\kappa\sin\psi
\ee
associated with the cross-sectional elements $\bfd$ and $\bft\times\bfd$ of the material frame, we arrive at an alternative representation
\be
\bfka-\kappa_0\bfd =(\kappa_1-\kappa_0)\bfd-\kappa_2\mskip2mu\bft\times\bfd
\label{bending_strain}
\ee
for the vectorial strain measure $\bfka-\kappa_0\bfd$. Finally, substituting \eqref{bending_strain} into the bending contribution to \eqref{E} we arrive at the bending energy
\be
\int_C \frac{1}{2}a((\kappa_1-\kappa_0)^2 + \kappa_2^2) 
\label{EZhou}
\ee
used by \citetalias{Haijun} to study DNA minicircles.

\subsection{Dimensionless parameters}
\label{scaling}

Consider a situation in which the loop made from a filament and fluid film are in a flat circular configuration, with radius $R>0$, and the filament is free of rotation about its midline, in which case the twist density obeys $\omega=0$. For any such configuration, the right-hand side of the total free-energy \eqref{E} specializes to
\be
E=\bigg ( (1 - R\kappa_0)^2 + \frac{c}{a}(R \omega_0)^2 + \frac{R^3 \sigma}{a} \bigg )\frac{\pi a}{R}.
\label{Eg}
\ee
Apart from showing that the total free-energy of a flat, circular configuration of radius $R$ scales with the lineal bending-energy $\pi a/R$ of $C$, \eqref{Eg} points to four potentially significant dimensionless parameters:
\be
\alpha = \frac{c}{a}>0, \qquad \beta = R \kappa_0\ge0, \qquad \gamma = R \omega_0, \qquad \eta = \frac{R^3\sigma}{a}\ge0.
\label{alphaetc}
\ee
The first of \eqref{alphaetc}, namely $\alpha$, represents the energy cost of twisting the filament relative to that of bending the filament. Following \cite{vogel}, we refer to $\alpha$ as the ``twist-to-bend ratio.'' The second and third of \eqref{alphaetc}, namely $\beta$ and $\gamma$, provide dimensionless measures of intrinsic curvature and intrinsic twist density; accordingly, we refer to them as the `dimensionless intrinsic curvature' and the `dimensionless intrinsic twist density.' In our study of stability and buckling, we will find that $\gamma$ appears only in combination with $\alpha$ through the product
\be
\mu=\alpha\gamma=\frac{Rc\omega_0}{a}.
\label{mu}
\ee
Finally, the last of \eqref{alphaetc}, namely $\eta$, is familiar from the work of \cite{chen} and measures the strength of the areal free-energy $\pi R^2\sigma$ of a flat circular film with radius $R$ and surface tension $\sigma$ relative to that of the lineal bending-energy $\pi a/R$ of a circular loop with radius $R$ and bending rigidity $a$; alternatively, $\eta$ can be thought of as the ratio of the magnitude of the force per unit length that the film exerts at a generic point of the circular loop of filament to the force per unit length, $a/R^3$, generated by the bending resistance of the filament. We refer to $\nu$ as the `dimensionless surface tension.'

\section{First and second variation conditions}
\label{variationconditions}

%


We assume that the system is isolated, from which it follows that its potential energy is simply the total free-energy \eqref{E}. Since \cite{aisa1,aisa2} provided detailed derivations of the first and second variations of the energy functional
\be
\bar{E}=\int_C \frac{1}{2}(a\kappa^2 + c\mskip1mu\omega^2)
+\int_S\sigma,
\label{barE}
\ee
we can take advantage of their results and focus on computing the first and second variations of the difference
\be
E_d=E-\bar{E}
=\int_C\Big(a\kappa_0\Big(\frac{\kappa_0}{2}-\kappa\cos\psi\Big)
+c\mskip1mu\omega_0\Big(\frac{\omega_0}{2}-\omega\Big)\Big).
\label{Ediff}
\ee
Consistent with the provision that $C$ is closed, our calculations rely on the periodicity of the Frenet frame $\{\bft,\bfp,\bfb\}$, the curvature $\kappa$, the torsion $\tau$, and the director $\bfd$. 


\subsection{First variation condition}
\label{firstvar}

Computing the first variation of \eqref{Ediff} gives
\be
\delta E_d
=-\int_C(a\kappa_0(\cos\psi\mskip2mu\delta\kappa-\kappa\sin\psi\mskip2mu\delta\psi)
+c\mskip1mu\omega_0\delta\omega).
\label{firstbending}
\ee
Let $\bfr$ be an arclength parameterization of $C$, so that $\bfr'=\bft$ and let $\bfu=\delta\bfr$ denote the variation of $\bfr$. Then, using (A.27) and (A.33) of \cite{aisa2} to replace $\delta\kappa$ and $\delta\tau$ in \eqref{firstbending} by relations involving $\bfu$, $\bfu'$, $\bfu''$, and $(\delta\psi)'$, we integrate by parts twice while invoking the periodicity of $\psi$, $\bfp$, $\bfu$, and $\bfu'$ to obtain an alternative version of \eqref{firstbending} which when combined with the expression for the first variation of \eqref{barE} derived in \S3.1 of \cite{aisa2} yields the first variation condition
\begin{multline}
\qquad
\delta E=\int_C[(a((\kappa-\kappa_0\cos\psi)\bfp)'
-c(\omega-\omega_0)\kappa\bfb-c(\kappa^{-1}\omega'\bfb)'-\bar\lambda\bft)'+\sigma\bft\times\bfn]\cdot \bfu
\\[4pt]
-\int_C(c\mskip1mu\omega'-a\kappa_0\kappa\sin\psi)\delta\psi-\int_S2\sigma H\bfn\cdot\bfu=0,
\qquad
\label{delE}
\end{multline}
where $\bar\lambda$ is a Lagrange multiplier needed to ensure the inextensibility of $C$. Standard localization arguments then yield the areal equilibrium condition
\be
H = 0
\label{meancurvaturezero}
\ee
expressing the normal component of force balance on $S$, a lineal equilibrium condition
%
%
\be
(a((\kappa-\kappa_0\cos\psi)\bfp)'
-c(\omega-\omega_0)\kappa\bfb-c(\kappa^{-1}\omega'\bfb)'  - \bar\lambda \bft)'  + \sigma \bft \times \bfn = \bold 0
\label{rodbalance1}
\ee
expressing force balance on $C$, and a second lineal equilibrium condition
\be
c\mskip1mu\omega'=a \kappa_0 \kappa \sin \psi
\label{rodbalance2}
\ee
expressing the tangential component of moment balance on $C$. These conditions are supplemented by the relation 
\be
\omega=\psi'+\tau
\label{omegarelbis}
\ee
for the twist density and by the constraint 
\be
|\bfr'|=|\bft|=1
\label{inextensibilityconstraint}
\ee
ensuring that $C$ is inextensible.

The areal equilibrium condition \eqref{meancurvaturezero} is familiar from the works of \cite{giomi}, \cite{chen}, and \cite{aisa1,aisa2}. If the intrinsic curvature $\kappa_0$ and the intrinsic twist density $\omega_0$ both vanish, the lineal equilibrium conditions \eqref{rodbalance1} and \eqref{rodbalance2} reduce to those derived by \cite{aisa1,aisa2}. Moreover, if we also require that the twisting rigidity $c$ vanishes, then \eqref{rodbalance1} reduces to the lineal equilibrium condition derived by \cite{chen} while \eqref{rodbalance2} is vacuous. Whereas the tangential component of \eqref{rodbalance1} yields a condition 
\be
\bar\lambda=\text{constant}-\frac{3}{2} a \kappa^2 - \frac{1}{2} c\mskip1mu\omega^2 + a \kappa_0 \kappa \cos \psi
\label{multplierrel}
\ee
for determining the reactive force $\bar\lambda$ needed to ensure the inextensibility of $C$. The remaining components
\be
\left.
\begin{aligned}
\kappa'' + \frac{1}{2} \kappa^3 + \kappa_0(\omega \sin \psi)' - \kappa \tau^2 + \kappa_0 \tau \omega \cos \psi + \frac{c \kappa \tau}{a} (\omega - \omega_0) - \frac{\kappa}{a}\Big(\chi - \frac{1}{2} c\omega^2\Big) - \frac{\sigma}{a} \sin \vartheta &= 0,
\\[4pt]
2 \kappa' \tau + \kappa \tau' + \kappa_0 \omega^2 \sin \psi - \kappa_0 \omega' \cos \psi - \frac{c}{a} (\omega - \omega_0) \kappa' - \frac{c}{a} \omega' \kappa + \frac{\sigma}{a} \cos \vartheta &= 0,
\end{aligned}
\,\right\}
\label{rodforcebalance}
\ee
of \eqref{rodbalance1} combine with \eqref{omegarelbis} to yield a coupled system for determining $\kappa$, $\tau$, and $\psi$ granted knowledge of the contact angle $\vartheta$. Also note that $\chi$ in \eqref{rodforcebalance}$_1$ is the constant on the right-hand side of \eqref{multplierrel}.

\subsection{Second variation condition}
\label{secondvar}

Computing the first variation of \eqref{firstbending} gives
\be
\delta^2 E_d
=-\int_C(a\kappa_0(\cos\psi\mskip2mu\delta^2\kappa
-2\sin\psi\mskip2mu\delta\kappa\mskip2mu\delta\psi
-\kappa\sin\psi\mskip2mu\delta^2\psi
-\kappa\cos\psi\mskip2mu(\delta\psi)^2)+c\mskip1mu\omega_0\delta^2\omega).
\label{secondbending}
\ee
Then, proceeding as in the treatment of \eqref{firstbending} and using (A.41) and (A.45) of \cite{aisa2} and \eqref{omegarel} to express $\delta^2\kappa$ and $\delta^2\omega$ in \eqref{secondbending} in terms of $\bfu$, $\bfu'$, $\bfu''$, and $(\delta^2\psi)'$, we integrate by parts twice to obtain an alternative version of \eqref{secondbending} which when combined with the expression for the second variation of \eqref{barE} derived in \S3.2 of \cite{aisa2} and the equilibrium conditions \eqref{meancurvaturezero}--\eqref{rodbalance2} yields the second variation condition
%
\begin{multline}
\qquad
\delta^2 E
=\int_C a ((1 - \kappa_0 \kappa^{-1}\cos \psi) |\bfu''|^2 + \kappa_0(\kappa^{-1} \cos \psi \left ( \bfp \cdot \bfu'' \right )^2 +2\sin \psi (\bfp \cdot \bfu'')\delta\psi + \kappa \cos \psi (\delta \psi)^2))
\\
+ \int_C c((\delta \omega)^2 - (\omega - \omega_0)\bft\cdot ( \bfu' \times \bfu'') - \omega' \delta (\kappa^{-1} \bfb) \cdot \bfu'')
+\int_C\lambda|\bfu'|^2 
\\
+\int_S\sigma((\divs\bfu)^2-\tr((\grads\bfu)^2)
+|(\grads\bfu) \bfn|^2)
\ge 0.
\qquad
\end{multline}

\section{Parametric reformulation}
\label{parametricrecasting}

\subsection{Parametrization}
\label{parametrization}

Suppose that surface $S$ admits a parametrization of the form
\be
S = \{\bfx \in \mathbb{R}^3, \bfx = R \bfxi ( r,\theta), \, 0 \le  r \le 1, \, 0 \le \theta \le 2 \pi\},  
\label{Sparametrization}
\ee
where $R$ is the radius of a circle with perimeter equal to the length $2\pi R$ of $C$, the dimensionless mapping $\bfxi$ is four-times continuously differentiable and injective, and the dimensionless radius $r$ and the azimuth $\theta$ provide polar coordinates on the closed unit disc (Figure~\ref{model}). Periodicity requires that
\be
\bfxi( r,\theta) = \bfxi( r,\theta + 2\pi), \qquad 0 \leq r\leq 1, \qquad 0 \leq \theta \leq 2 \pi,
\ee
and so on for all relevant higher derivatives of $\bfxi$. With \eqref{Sparametrization}, the assumption \eqref{coincidence} that $\partial S$ and $C$ coincide induces a parametrization
\be
C = \{\bfx \in \mathbb{R}^3, \bfx = R \bfxi (1,\theta), \, 0 \le \theta \le 2 \pi\}
\label{Cparametrization}
\ee
of $C$ and the constraint of inextensibility takes the form
\be
|\bfxi_\theta(1,\theta)|=1,
\qquad
0\le\theta\le2\pi.
\label{xiconstraint}
\ee
Additionally, we define $\Psi$ by
\be
\Psi(\theta)=\psi(R\theta),
\qquad
0\le\theta\le2\pi.
\label{Psi}
\ee

From this point onward, we treat all quantities defined on $C$ as functions of the polar angle $\theta$. Using the parametrization \eqref{Cparametrization}, we accordingly find that the curvature $\kappa$, torsion $\tau$, and twist density $\omega$ of $C$ are given by
\be
\kappa=\frac{|\xitt|}{R},
\qquad 
\tau=\frac{(\xit\stimes\xitt)\sdot\xittt}{R|\xitt|^2},
\qquad\text{and}\qquad
\omega=\frac{\Psi_\theta}{R}
+\frac{(\xit\stimes\xitt)\sdot\xittt}{R|\xitt|^2}
.
\label{kappatauomegaparrep}
\ee
Similarly, we find that the tangent, normal, and binormal elements $\bft$, $\bfp$, and $\bfb$ of the Frenet frame of $C$ are given by
\be
\bft = \xit,
\qquad 
\bfp = \frac{\xitt}{|\xitt|},
\qquad\text{and}\qquad
\bfb = \frac{\xit\stimes\xitt}{|\xitt|}.
\label{tpbrep}
\ee
By \eqref{FSrlns}$_1$, \eqref{kappatauomegaparrep}$_1$, and \eqref{tpbrep}$_2$, the vector curvature  $\bfka$ of $C$ is given by
\be
\bfka=\frac{\xitt}{R}.
\label{bfkarep}
\ee
Additionally, by \eqref{drepresentation}, \eqref{Psi}, and \eqref{tpbrep}$_{2,3}$, the director $\bfd$ is given by
\be
\bfd=\frac{\cos\Psi\mskip2mu\xitt+\sin\Psi\mskip2mu\xit\stimes\xitt}{|\xitt|}.
\label{drep2}
\ee
Since $\{\bft,\bfp,\bfb\}$, $\kappa$, $\tau$, $\omega$, $\bfka$, and $\bfd$ are defined only on $C$, \eqref{kappatauomegaparrep}--\eqref{drep2} make sense only if the quantities on their right-hand sides are evaluated at $r=1$. We avoid cumbersome notation by leaving such evaluations implicit unless, as in the case of the constraint \eqref{xiconstraint}, there would otherwise be potential for confusion. 

\subsection{Dimensionless free-energy}
\label{dimensionlessfreeenergy}

Using \eqref{Cparametrization}, \eqref{xiconstraint}, \eqref{kappatauomegaparrep}$_3$, \eqref{bfkarep}, and \eqref{drep2} in the lineal contribution to the total free-energy \eqref{E}, we find that
\begin{multline}
\int_{C}\frac{1}{2}(a|\bfka-\kappa_0\bfd|^2+c(\omega-\omega_0)^2)
\\[4pt]
=\frac{a}{R}
\lint\frac{1}{2}\bigg[(|\xitt|^2-2\beta|\xitt|\cos\Psi+\beta^2)+\alpha\bigg(\Psi_\theta+\frac{(\xit\stimes\xitt)\sdot\xittt}{|\xitt|^2}
-\gamma\bigg)^{\!\!2}\,\bigg]\dtheta,
\label{Fl}
\end{multline}
where $\alpha=c/a>0$, $\beta=R\kappa_0\ge0$, and $\gamma=R\omega_0$ denote the twist-to-bend ratio, the dimensionless intrinsic curvature, and the dimensionless intrinsic twist density defined, respectively, by \eqref{alphaetc}$_1$, \eqref{alphaetc}$_2$, and \eqref{alphaetc}$_3$.
%
%
%
Moreover, since an element of $S$ spanned by $\xir\text{d}r$ and $\xit\text{d}\theta$ has area $|\xir\dr\stimes\xit\dtheta|=|\xir\stimes\xit|\drdtheta$, the areal contribution to the total free-energy \eqref{E} has the simple form
\be
\int_{S}\sigma=\frac{a}{R}\aint\eta|\xir\times\xit|\drdtheta,
\label{Fa}
\ee
where $\eta=R^3\sigma/a\ge0$ denotes the dimensionless surface tension defined by \eqref{alphaetc}$_4$.

Like the net potential-energy \eqref{Eg} of a flat, circular configuration with vanishing twist density, the lineal and areal contributions \eqref{Fl} and \eqref{Fa} to the total free-energy scale with $a/R$. It is therefore convenient to work with the dimensionless total free-energy
\begin{multline}
\Phi=\frac{E}{a/R}
=\lint\frac{1}{2}\bigg(|\xitt|^2-2\beta|\xitt|\cos\Psi+\beta^2+\alpha\bigg(\Psi_\theta+\frac{(\xit\stimes\xitt)\sdot\xittt}{|\xitt|^2}
-\gamma\bigg)^{\!\!2}\,\bigg)\dtheta
\\[4pt]
+\aint\eta|\xir\times\xit|\drdtheta.
\label{Phi}
\end{multline}

\subsection{Dimensionless first variation condition}
\label{dimensionlessfirstvar}

For the dimensionless total free-energy \eqref{Phi}, the parametric dimensionless counterpart of the first variation condition \eqref{delE} has the form
\begin{multline}
\qquad
\delta\Phi=\lint\bigg\{\bigg[\bigg(\xitt-\beta\cos\Psi\frac{\xitt}{|\xitt|}\bigg)_{\subthet}
-\alpha\bigg(\Psi_\theta+\frac{(\xit\stimes\xitt)\sdot\xittt}{|\xitt|^2}
-\gamma\bigg)\frac{\xitt\times\xittt}{|\xitt|^2}
\\[4pt]
-\alpha\bigg[\bigg(\Psi_\theta+\frac{(\xit\stimes\xitt)\sdot\xittt}{|\xitt|^2}
\bigg)_{\subthet}\frac{\xit\stimes\xitt}{|\xitt|^2}\bigg]_{\mskip-2mu\theta}-\lambda\xit\bigg]_{\mskip-2mu\theta}+\eta\xit\times\bfn\bigg\}\cdot\bfv\dtheta
\\[4pt]
-\lint\bigg(
\alpha\bigg(\Psi_\theta+\frac{(\xit\stimes\xitt)\sdot\xittt}{|\xitt|^2}
\bigg)_{\subthet\mskip2mu}-\beta|\xitt|\sin\Psi\bigg)\delta\Psi\dtheta
\\[4pt]
-\aint\eta(\bfxi_\theta \times \bfn_ r + \bfn_\theta \times \bfxi_ r) \cdot \bfv\drdtheta=0,
\qquad
\label{delPhi}
\end{multline} 
where $\bfv=\delta\bfxi$ is the dimensionless variation of $\bfxi$ and
\be
\lambda=\frac{R^2\bar\lambda}{a},
\label{lambda}
\ee
is a dimensionless Lagrange multiplier required to ensure satisfaction of the constraint \eqref{xiconstraint}.

Localizing \eqref{delPhi} yields dimensionless counterparts of the equilibrium conditions \eqref{meancurvaturezero}--\eqref{rodbalance2}. \cite{chen} show that the parenthesized quantity in the areal contribution to \eqref{delPhi} obeys $\bfP(\bfxi_\theta \stimes \bfn_ r + \bfn_\theta \stimes \bfxi_ r)=\bf0$ with $\bfP=\bfI-\bfn\otimes\bfn$,
%
%
Thus, choosing $\bfv$ to be compactly supported on a proper subset of the unit disk and using the requirement that $\delta\Psi$ must vanish in equilibrium yields the areal force balance
\be
\bfn\cdot(\bfxi_\theta \stimes \bfn_ r + \bfn_\theta \stimes \bfxi_ r)=0.
\label{areal}
\ee
Further, choosing $\bfv$ to be compactly supported on a proper subset of the boundary of the unit disk and using the requirement that $\delta\Psi$ must vanish in equilibrium yields the lineal force balance
\begin{multline}
\bigg(\xittt
-\alpha\bigg(\frac{(\xit\stimes\xitt)\sdot\xittt}{|\xitt|^2} + \Psi_\theta-\gamma\bigg)\frac{\xitt\times\xittt}{|\xitt|^2}
\\[4pt]
-\beta\bigg(\frac{\cos\Psi\mskip2mu\xitt+\sin\Psi\mskip2mu\xit\stimes\xitt}{|\xitt|}\bigg)_{\subthet}-\lambda\xit\bigg)_{\subthet} 
+\eta\mskip1mu\xit\times\bfn=\boldsymbol{0}.
\label{lineal1}
\end{multline}
Finally choosing $\delta\Psi$ to be compactly supported on a proper subset of the boundary of the unit disk and using the requirement that $\delta\Psi$ must vanish in equilibrium yields the tangential component
\be
\alpha\bigg(\Psi_\theta
+\frac{(\xit\stimes\xitt)\sdot\xittt}{|\xitt|^2}
\bigg)_{\subthet} 
=\beta|\xitt|\sin\Psi
\label{lineal2}
\ee
of the  lineal moment balance. The equilibrium conditions \eqref{areal}--\eqref{lineal2} must be supplemented by the constraint \eqref{xiconstraint} of inextensibility, which we repeat here for completeness:
\be
|\bfxi_\theta(1,\theta)|=1,
\qquad
0\le\theta\le2\pi.
\label{xiconstraintbis}
\ee
If the filament is free of intrinsic curvature and intrinsic twist, so that, by \eqref{alphaetc}$_{2,3}$, the dimensionless intrinsic curvature $\beta$ and the dimensionless intrinsic twist density $\gamma$ both vanish, \eqref{lineal1} and \eqref{lineal2} reduce to (5.13) and (5.12), respectively, of \cite{aisa1}.  

\subsection{Dimensionless second variation condition}
\label{dimensionlesssecondvar}

For the dimensionless total free-energy \eqref{Phi}, the parametric dimensionless counterpart of the second variation condition \eqref{delE} has the form
\begin{multline}
\delta^2 \Phi =
\lint\bigg[\bigg(1-\frac{\beta\cos\Psi}{|\xitt|}\bigg)|\vtt|^2
+\beta|\xitt|\cos\Psi(\delta\Psi)^2
+\frac{\beta\cos\Psi(\xitt\sdot\vtt)^2}{|\xitt|^3}+\frac{2\beta\sin\Psi(\xitt\sdot\vtt)\delta\Psi}{|\xitt|}
\\[4pt]
-\alpha\bigg(\Psi_\theta+\frac{(\xit\stimes\xitt)\sdot\xittt}{|\xitt|^2}
- \gamma \bigg)\xit\cdot(\vt\times\vtt)
+\lambda|\vt|^2
+\alpha\bigg[(\delta\Psi)_\theta+\delta\bigg(\frac{(\xit\stimes\xitt)\sdot\xittt}{|\xitt|^2}\bigg)\bigg]^2  
\\[4pt]
+\alpha \bigg(\Psi_\theta+\frac{(\xit\stimes\xitt)\sdot\xittt}{|\xitt|^2}\bigg)_{\subthet}\bigg( \frac{(\xit\stimes\xitt)\sdot\vt}{|\xitt|^2}(\xit\sdot\vtt)
+\frac{2(\xitt\sdot\vtt)((\xit\stimes\xitt)\cdot\vtt)}{|\xitt|^4}\bigg)\bigg]\dtheta
\\[4pt]
+\aint\eta\bigg(\frac{|\bfP(\bfv_ r\stimes\bfxi_\theta+\bfxi_r\stimes\bfv_\theta)|^2}
{|(\bfxi_r\stimes\bfxi_\theta|}+2\bfn\sdot(\bfv_r\stimes\bfv_\theta)\bigg)\drdtheta
\ge 0.
\label{stabilitycondition}
\end{multline}
If the filament is free of intrinsic curvature and intrinsic twist, so that, by \eqref{alphaetc}$_{2,3}$, the dimensionless intrinsic curvature $\beta$ and the dimensionless intrinsic twist density $\gamma$ both vanish, \eqref{stabilitycondition} reduces to (39) of \cite{aisa2}.

\section{Stability analysis}
\label{stability}

We now study the stability of a flat circular configuration in which the cross sections of the filament are not rotated about its midline. To describe such a configuration, $\bfxi$ must parametrize a flat disk of unit radius and therefore be of the form
\be
\bfxi(r,\theta)=r\bfe(\theta),
\qquad
0\le r\le 1,
\qquad
0\le\theta\le2\pi,
\label{trvlsln1}
\ee
where $\bfe$ denotes the radial basis vector, and $\Psi$ must obey
\be
\Psi(\theta)=0,
\qquad
0\le\theta\le2\pi.
\label{trvlsln2}
\ee

For the choices \eqref{trvlsln1} and \eqref{trvlsln2} of $\bfxi$ and $\Psi$, the areal force balance \eqref{areal} is satisfied identically, as are the components of the projection of the lineal force balance \eqref{lineal1} onto the plane orthogonal to the tangent of the midline, the lineal moment balance \eqref{lineal2}, and the constraint \eqref{xiconstraintbis} of inextensbility. The only remaining condition, namely the component of the lineal force balance \eqref{lineal1} parallel to the tangent of the midline, leads to the conclusion that the dimensionless Lagrange multiplier $\lambda$ is a constant:
\be
\lambda=\beta-(1+\eta).
\label{trvlsln3}
\ee
For brevity, we refer to the combination of $\bfxi$, $\Psi$, and $\lambda$ given by \eqref{trvlsln1}--\eqref{trvlsln3} as the `trivial solution' of \eqref{areal}--\eqref{xiconstraintbis}. Moreover, we refer to the family of such solutions determined by all combinations of the dimensionless intrinsic curvature $\beta\ge0$ and the dimensionless surface tension $\eta>0$ as the `trivial solution branch.'

We consider a perturbation 
\be
\bfxi(r,\theta)=r\bfe(\theta)+\bfv(r,\theta),
\qquad
0\le r\le 1,
\qquad
0\le\theta\le2\pi,
\label{xiplusv}
\ee
of $\bfxi$, with $|\bfv|\ll1$. Introducing  the azimuthal basis vector $\bfe^{\sperp}$, we define the azimuthal, radial, and transverse components of $\bfv$ according to:
\be
u=\bfe^{\sperp}\cdot\bfv,
\qquad
v=\bfe\cdot\bfv,
\qquad\text{and}\qquad
w=(\bfe\times\bfe^{\sperp})\cdot\bfv.
\label{vcmpnnts}
\ee
Then, since $\bfe^{\sperp}_\theta=\bfe_{\theta\theta}=-\bfe$, \eqref{xiconstraint} requires that, to most significant order in $\bfv$,
\be
\bfv_\theta(1,\theta)\cdot\bfe^{\sperp}(\theta)
=u_\theta(1,\theta)+v(1,\theta)=0,
\qquad
0\le\theta\le2\pi.
\label{vconstraint}
\ee
Furthermore, using \eqref{xiplusv}--\eqref{vconstraint} in \eqref{stabilitycondition}, we find that the second variation about the trivial solution takes the form
\begin{multline}
\lint(\alpha(w_\theta+ w_{\theta\theta\theta}+(\delta\Psi)_\theta)^2
+2\mu(v+v_{\theta\theta})w_\theta+(1-\beta)w_{\theta\theta}^2
-(1+\eta-\beta)w_\theta^2\\[4pt]
+(v+v_{\theta\theta})^2 + \beta (\delta\Psi)^2+ \eta (v^2 - v_\theta^2))\dtheta
+\aint\eta\bigg(rw_r^2+\frac{1}{r}w_\theta^2\bigg)\drdtheta\ge0,
\label{secondvariationcomponent}
\end{multline}
where, recalling \eqref{mu}, $\mu$ is the product of the twist-to-bend ratio $\alpha=c/a>0$ and the dimensionless intrinsic twist density $\gamma=R\omega_0$, namely $\mu=\alpha\gamma=Rc\omega_0/a$. 

%

From \eqref{secondvariationcomponent}, it is evident that the azimuthal component of $\bfv$ has no influence on the stability of the trivial solution. To extract further information from \eqref{secondvariationcomponent}, we therefore introduce Fourier expansions 
\be
\left.
\begin{aligned}
v(r,\theta)
&=\sum_{n=0}^{\infty}(a_n(r)\cos n\theta+b_n(r)\sin n\theta),
\\[4pt]
w(r,\theta)
&=\sum_{n=0}^{\infty}(c_n(r)\cos n\theta+d_n(r)\sin n\theta),
\end{aligned}
\,\right\}
\qquad
0\le r\le1,
\qquad
0\le\theta\le2\pi,
\label{vwFourier}
\ee
for the remaining components of $\bfv$ along with a Fourier expansion,
\be
\delta\Psi(\theta)
=\sum_{n=0}^{\infty}(e_n\cos n\theta+f_n\sin n\theta),
\qquad
0\le\theta\le2\pi,
\label{PsiFourier}
\ee
for $\delta\Psi$. Using \eqref{vwFourier} and \eqref{PsiFourier} in \eqref{secondvariationcomponent}, we emulate calculations of \cite{chen} and \cite{aisa2} to arrive at the stability condition
\be
2\beta{e}^2_0
+\sum_{n=1}^\infty S_n
+2(1+\eta){a}^2_0(1)
+2\eta\int_0^1\bigg|\Dr{c_0}\bigg|^2r\dr 
+\eta\sum\limits_{n=1}^\infty\int_0^1
\bigg(\bigg|\Dr{c_n}-\frac{nc_n}{r}\bigg|^2
+\bigg|\Dr{d_n}-\frac{nd_n}{r}\bigg|^2\bigg)r\dr
\ge0,
\label{tss}
\ee
%
%
where the summand $S_n$ is defined by
%
%
%
%
\begin{multline}
S_n = (n^2-1)(n^2-1-\nu)(a^2_n(1) + b^2_n(1)) 
+ (\alpha n^2+\beta)(e^2_n + f^2_n) 
\\[4pt]
+ ((1 - \beta)n^4 - (1 + \eta - \beta)n^2 + \alpha n^2 (n^2 - 1)^2 + \eta n) (c^2_n(1)+d^2_n(1)) 
\\[4pt]
- 2 n(n^2-1)(\alpha n (c_n(1)e_n + d_n(1)f_n) + \mu (a_n(1)d_n(1)-b_n(1)c_n(1))).
\qquad
\end{multline}
Since the dimensionless intrinsic curvature $\beta$ and the dimensionless surface tension $\eta$ are nonnegative, the inequality \eqref{tss} holds and the trivial solution is stable if
\be
S_n\ge0
\qquad
\text{for all}
\qquad
n\ge1.
\label{Scond}
\ee
%

On defining
\be
\left. 
\begin{array}{c}
A = (n^2-1)(n^2-1-\eta), 
\qquad B = (1 - \beta)n^4 - (1 + \eta - \beta)n^2 + \alpha n^2 (n^2 - 1)^2 + \eta n,  
\cr\noalign{\vskip6pt}
C = \alpha n^2 + \beta, 
\qquad D = \mu n (n^2-1), 
\qquad\text{and}\qquad E = \alpha n^2 (n^2-1),
\end{array} 
\right\}
\label{Coefficients}
\ee
it can be shown that a condition necessary for \eqref{Scond} to be satisfied is that the symmetric matrix
\be
\left[
\begin{array}{cccccc}
\phantom{-}A & \phantom{-}0 & \phantom{-}0 & -D & \phantom{-}0 & \phantom{-}0 
\\ 
\phantom{-}0 & \phantom{-}A & \phantom{-}D & \phantom{-}0 & \phantom{-}0 & \phantom{-}0 
\\ 
\phantom{-}0 & \phantom{-}D & \phantom{-}B & \phantom{-}0 & -E & \phantom{-}0 \\ 
-D & \phantom{-}0  & \phantom{-}0 & \phantom{-}B & \phantom{-}0 & -E 
\\ 
\phantom{-}0 & \phantom{-}0 &  -E & \phantom{-}0  & \phantom{-}C & \phantom{-}0 
\\ 
\phantom{-}0 &  \phantom{-}0 & \phantom{-}0 & -E & \phantom{-}0 & \phantom{-}C 
\end{array}
\right]
\label{symmatrix}
\ee
be semipositive definite. Applying Corollary 7.2.4 of \cite{Horn}, which states that a real symmetric matrix is positive semidefinite if and only if all the coefficients of its characteristic polynomial alternate in sign, to \eqref{symmatrix} yields
%
\be
A \ge 0
\qquad\text{and}\qquad 
ABC \ge C D^2 + A E^2.
\label{Inequalities}
\ee
For the particular choice $n=1$, which corresponds to a rigid rotation of the system consisting of the loop and the spanning film, the definitions \eqref{Coefficients} specialize to give $A=B=D=E=0$ and both of \eqref{Inequalities} hold as identities. Accordingly, we restrict attention to values of $n\ge2$ from this point onward.

Introducing the dimensionless combination
\be
k_n
=\frac{(\alpha+\beta)\beta}{\alpha n^2+\beta}\ge0
\label{ksubn}
\ee
of the twist-to-bend ratio $\alpha$, the dimensionless intrinsic curvature $\beta$, and the mode number $n$, we next show that the stability conditions \eqref{Inequalities} are equivalent to the requirement that the three inequalities
%
%
%
\be
\eta\le\frac{n+1}{2}\left(2n-1-nk_n
-\sqrt{{\displaystyle(1 - nk_n)^2} + \frac{4\mu^2n}{n+1}}\,\right),
\qquad
\mu^2\le(n^2-1)(1-k_n),
\qquad
k_n\le1
\label{stabilitycriteria}
\ee
hold for each $n\ge2$. To show that \eqref{stabilitycriteria} are consequences of \eqref{Inequalities}, we use \eqref{Coefficients} to replace $A$, $B$, $C$, $D$, and $E$ in \eqref{Inequalities}. Taking advantage of the definition \eqref{ksubn} of $k_n$ to simplify the resulting expressions, we then arrive at two inequalities
\be
\eta\le n^2-1
\qquad\text{and}\qquad
q_n(\eta)\ge0,
\qquad
n\ge2,
\label{stability1} 
\ee
where $q_n$ is the quadratic defined by
\be
q_n(\eta) = \eta^2 - (n+1)(2n -1 - nk_n) \eta + n(n+1)( (n^2-1)(1-k_n) - \mu^2).
\label{quadratic}
\ee
Since $q_n(\eta)|_{\eta=n^2-1}=-n(n+1)\mu^2\le0$ for each $n\ge2$, the roots
\be
\eta_\pm=\frac{n+1}{2}\left(2n-1-nk_n
\pm\sqrt{{\displaystyle(1 - nk_n)^2} + \frac{4\mu^2n}{n+1}} \,\right)
\label{roots}
\ee
of the quadratic equation $q_n(\eta)=0$ must satisfy
\be
\eta_{-} \le n^2 - 1 \le \eta_{+},
\qquad
n\ge2.
\label{stability2}
\ee
Combining \eqref{stability1}$_1$ and \eqref{stability2}, we thus arrive at \eqref{stabilitycriteria}$_1$. Next, since $\eta$ is nonnegative, $q_n(0)$ must be nonnegative or, equivalently, $\mu$ must obey
\be
\mu^2\le(n^2-1)(1-k_n),
\qquad
n\ge2.
\label{musquared}
\ee
Moreover, on combining \eqref{stabilitycriteria}$_1$ and \eqref{musquared} while bearing in mind that $\mu^2$ is nonnegative, we arrive at both \eqref{stabilitycriteria}$_2$ and \eqref{stabilitycriteria}$_3$.  This confirms that \eqref{stabilitycriteria} follows from \eqref{Inequalities}. To establish the converse assertion, namely that \eqref{Inequalities} follow from \eqref{stabilitycriteria}, we observe that, since both conditions in \eqref{stability1} follow from \eqref{Inequalities}, we may reverse the steps leading from \eqref{Inequalities} to \eqref{stability1} to obtain \eqref{Inequalities}.

\begin{figure}[!t]
\begin{center}
\includegraphics[width = 90mm,height=75mm]{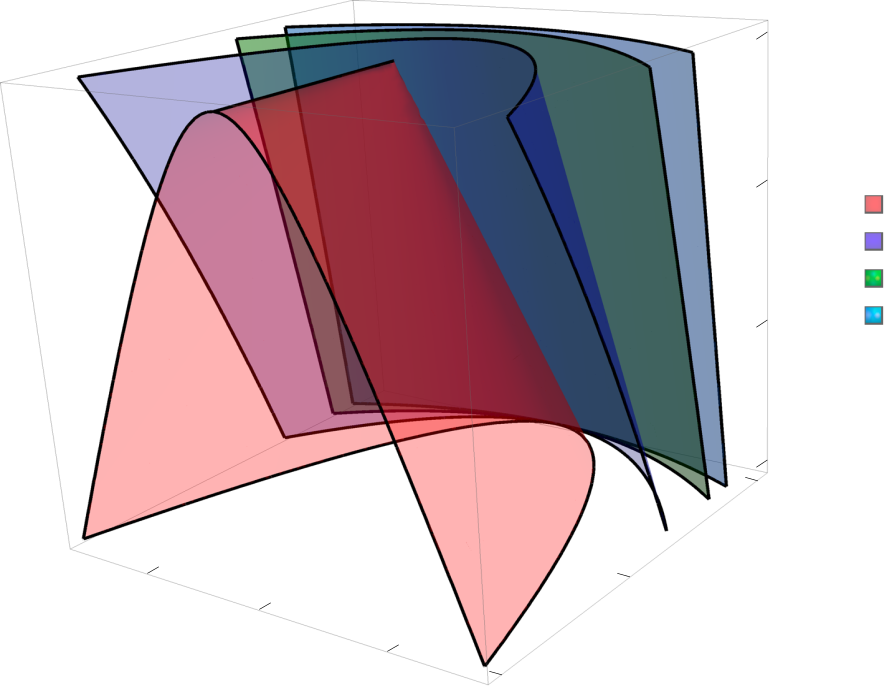}
\put(-224,25){$-1$}
\put(-187,15){0}
\put(-150,2){1}
\put(-112,-5){0}
\put(-75,25){0.5}
\put(-40,53){1}
\put(-33,68){0}
\put(-33,112){1}
\put(-33,155){2}
\put(-33,200){3}
\put(-195,5){$\mu$}
\put(-60,20){$k_n$}
\put(-25,130){$\nu$}
\put(0,148){\small $n=2$}
\put(0,137){\small $n=3$}
\put(0,126){\small $n=4$}
\put(0,114){\small $n=5$}
\caption{Stability domain in the presence of both intrinsic curvature and twist for the modes $n=2$, $3$, $4$, and $5$.}
\label{StabilityDomain}
\end{center}
\end{figure}
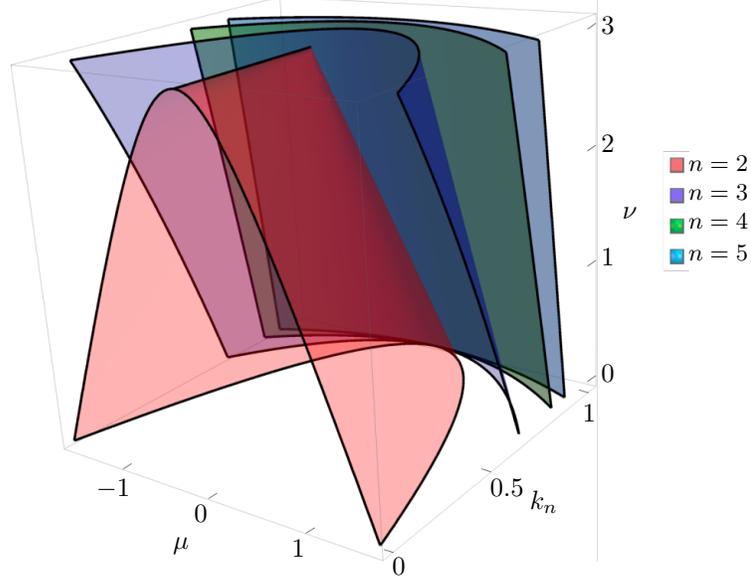
The conditions \eqref{stabilitycriteria} are sufficient to ensure that the trivial solution is stable. That these conditions are also necessary for the trivial solution to be stable becomes evident on using
\be
\left.
\begin{aligned}
v(r,\theta)
&=r^n(a_n(1)\cos n\theta+b_n(1)\sin n\theta),
\\[4pt]
w(r,\theta)
&=r^n(c_n(1)\cos n\theta+d_n(1)\sin n\theta),
\\[4pt]
\delta\Psi(\theta)
&=e_n\cos n\theta+f_n\sin n\theta,
\end{aligned}
\,\right\}
\qquad
0\le r\le1,
\qquad
0\le\theta\le2\pi,
\qquad
n\ge2,
\label{particularchoice}
\ee
in \eqref{secondvariationcomponent}, which leads to the conclusion that $S_n$ must satisfy $S_n\ge0$ for all $n\ge2.$

For each mode $n\ge2$, those combinations of the dimensionless surface tension $\eta$, the product $\mu=\alpha\gamma$ of the twist-to-bend ratio $\alpha$ and the dimensionless intrinsic twist density $\gamma$, and the composite parameter $k_n$ defined in \eqref{ksubn} consistent with \eqref{stabilitycriteria} determine a convex three-dimensional region (Figure~\ref{StabilityDomain})
\be
\Pi_n=\bigg\{(\eta,\mu,k_n):
0\le\eta\le\frac{3}{2}\bigg(3-nk_n-\sqrt{\displaystyle(1-nk_n)^2+\frac{8\mu^2}{3}}\,\bigg),\mu^2\le3(1-k_n),0\le k_n\le1\bigg\}.
\label{Rsubn}
\ee
Noting that $\Pi_n$ is a proper subregion of $\Pi_{n+1}$ for each $n\ge2$, we identify $n=2$ as the least stable mode number. Recalling from \eqref{alphaetc}$_4$ and \eqref{mu} that $\eta=R^3\sigma/a$ and $\mu=\alpha\gamma=Rc\omega_0/a$, it is easy to conceive of ways to tune $R$, $a$, $c$, $\omega$, and $\sigma$ to ensure that, for any choice of $n\ge2$, $\eta$ and $\mu$ belong to $\Pi_n$ for any particular choice of $0\le k_n\le1$. Since the intrinsic curvature $\kappa_0$ appears only in the composite parameter $k_n$ defined in \eqref{ksubn}, its potential influence is less transparent. From \eqref{ksubn}, it follows that, for any mode $n$ satisfying $n\ge2$, $k_n$ increases monotonically as a function of $\beta$ for any choice of $\alpha>0$ but decreases monotonically as a function of $\alpha$ for any choice of $\beta>0$. Recalling from \eqref{alphaetc}$_{1,2}$ that $\alpha=c/a>0$ and $\beta=R\kappa_0\ge0$, we arrive at the intuitively appealing conclusion that the potentially destabilizing influence of $\kappa_0$ can be countered by increasing the twisting rigidity $c$ relative to the bending rigidity $a$.

\section{Buckling analysis}
\label{buckling}

\subsection{General results}

The boundary-value problem \eqref{areal}--\eqref{xiconstraintbis} has a nontrivial solution branch that bifurcates from the trivial solution branch only if its linearization about a solution of the form \eqref{trvlsln1}--\eqref{trvlsln3} has a nonvanishing solution. To explore the existence of such bifurcation solutions, we take $\bfxi$ to be given by \eqref{xiplusv}--\eqref{vcmpnnts} and $\lambda$ to be such that
\be
\lambda (\theta) = \beta - (1+\eta) + \epsilon(\theta),
\qquad
0\le\theta\le2\pi.
\ee
With the objective of linearizing \eqref{areal}--\eqref{xiconstraint}, we assume that $|\bfv|\ll1$, $|\Psi|\ll1$, and $|\epsilon|\ll1$. From the areal force balance \eqref{areal}, we find that the transverse component $w$ of the incremental correction $\bfv$ to $\bfxi$ must satisfy the Laplace equation
\be
w_{rr} + \frac{1}{r} w_r +  \frac{1}{r^2} w_{\theta\theta} = 0
\label{diskshape}
\ee
on the unit disk. Moreover, from the lineal force balance \eqref{lineal1}, the linearized consequence \eqref{vconstraint} of the constraint \eqref{xiconstraint}, we find that radial and transverse components $v$ and $w$ of $\bfv$, the angle $\Psi$, and the incremental correction $\epsilon$ to the dimensionless Lagrange multiplier $\lambda$ must satisfy three boundary conditions, 
\be
\left.
\begin{array}{c}
(1 - \beta) w_{\theta\theta\theta\theta} + (\eta - \beta + 1) w_{\theta\theta} - \mu (v_{\theta\theta\theta} + v_{\theta}) + \eta w_r - \beta (\Psi_{\theta\theta} + \Psi) = 0,
\cr\noalign{\vskip8pt}
v_{\theta\theta\theta\theta} + (\eta + \beta - 1) v_{\theta\theta} + (\eta + \beta - 2) v + \epsilon = 0,
\cr\noalign{\vskip10pt}
(3 - \beta) (v_{\theta\theta\theta}+ v_{\theta}) + \mu (w_{\theta\theta\theta\theta} + w_{\theta\theta}) - \epsilon_{\theta}  = 0,
\end{array}
\,\right\}
\label{rodshape1}
\ee
on the unit circle. Finally, from the lineal moment balance \eqref{lineal2}, we find that $w$ and $\Psi$ must satisfy the additional boundary condition
\be
\alpha( w_{\theta\theta\theta\theta} +w_{\theta\theta}) + \alpha\Psi_{\theta\theta} - \beta\Psi = 0
\label{rodshape2}
\ee
on the unit circle.

Proceeding much as in Section~\ref{stability}, we assume that $v$ and $w$ have Fourier expansions of the form \eqref{vwFourier}, that $\Psi$ has a Fourier expansion with structure identical to that used previously to represent $\delta\Psi$, so that
\be
\Psi(\theta)
=\sum_{n=0}^{\infty}(e_n\cos n\theta+f_n\sin n\theta),
\qquad
0\le\theta\le2\pi.
\label{PsiFourierbis}
\ee
and that $\epsilon$ has a Fourier expansion of the form
\be
\epsilon(\theta) =\sum_{n=0}^{\infty}(p_n\cos n\theta+q_n\sin n\theta),
\qquad
0\le\theta\le2\pi.
\label{eFourier}
\ee
To ensure that \eqref{diskshape} is satisfied, the expansion \eqref{vwFourier}$_2$ for $w$ must take the form
\be
w(r,\theta)=\sum_{n=0}^{\infty}r^n(c_n(1)\cos n\theta+d_n(1)\sin n\theta),
\qquad
0\le r\le1,
\qquad
0\le\theta\le2\pi.
\label{wsol}
\ee
%
%

Using \eqref{vwFourier}$_1$ and \eqref{PsiFourierbis}--\eqref{wsol} in the boundary conditions \eqref{rodshape1} and \eqref{rodshape2}, we obtain a homogeneous system of eight equations for the eight unknowns $a_n(1), b_n(1), c_n(1), d_n(1), e_n, f_n, p_n$, and $q_n$.  We then eliminate $e_n, f_n, p_n$, and $q_n$ to arrive at a pair,
\be
n (n-1)
\begin{bmatrix}
(n+1) (n^2 - 1 - \eta) & -n(n+1)\mu  \\ 
-(n+1) \mu & n(n+1)(1-k_n) - \eta  \\ 
\end{bmatrix}
\begin{bmatrix}
a_n(1)  \\ 
d_n(1) \\ 
\end{bmatrix}
= 
\begin{bmatrix}
0 \\ 
0 \\ 
\end{bmatrix}
\label{ad}
\ee
and
\be
n (n-1)
\begin{bmatrix}
(n+1) (n^2 - 1 - \eta) & n(n+1)\mu  \\ 
(n+1) \mu & n(n+1)(1-k_n) - \eta  \\ 
\end{bmatrix}
\begin{bmatrix}
b_n(1) \\ 
c_n(1) \\ 
\end{bmatrix}
= 
\begin{bmatrix}
0 \\ 
0 \\ 
\end{bmatrix}
,
\label{bc}
\ee
of homogeneous systems for the remaining unknowns $a_n(1), b_n(1), c_n(1)$, and $d_n(1)$. 
For \eqref{ad} and \eqref{bc} to possess nontrivial solutions, the common determinant of their coefficient matrices must vanish. With reference to the definition \eqref{quadratic} of $q_n$, we are therefore led to a single solvability condition:
\be
n(n^2-1)q_n(\eta) = 0.
\label{solvcond}
\ee
In \eqref{solvcond}, the choices $n=0$ and $n=1$ describe rigid body translations and rotations, respectively, and are thus of no physical interest. For $n\ge2$, \eqref{solvcond} holds if $q_n(\eta)=0$. With reference to the stability condition \eqref{stabilitycriteria}, this requires that $\eta$, $\mu$, and $k_n$ satisfy
\be
\eta=\frac{n+1}{2} \left( 2n-1-nk_n - \sqrt{{\displaystyle(1 - nk_n)^2} + \frac{4\mu^2n}{n+1}}\,\right), \quad\,\, \mu^2 \le (n^2 - 1) (1 - k_n), \quad\,\, k_n \le 1,
\label{etasolution}
\ee
for each $n\ge2$.

For buckling to occur, \eqref{etasolution} must hold. This requirement can be specialized to recover some previously published results: 
\begin{itemize}
\item Situations in which the loop is not spanned by a fluid film can be realized by setting the surface tension $\sigma$ to zero. For $\sigma=0$ (or, equivalently, by \eqref{alphaetc}$_4$, $\nu=0$), \eqref{etasolution}$_1$ reduces to 
\be 
\mu^2=(n^2-1)(1-k_n),
\label{noten}
\ee
from which it follows that \eqref{etasolution}$_2$ holds automatically. 
Together, \eqref{etasolution}$_3$ and \eqref{noten} are equivalent to those derived by \citetalias{Haijun} for the buckling of a loop of length $2\pi R$ made from a filament with bending rigidity $a>0$, twisting rigidity $c>0$, intrinsic curvature $\kappa_0\ge0$ and intrinsic twist density $\omega_0$. This specialization shows that, in the absence of a spanning film, the presence of intrinsic curvature leads to buckling at values of $\mu=\alpha\gamma=Rc\omega_0/a$ below the critical value for a loop made from a filament with intrinsic twist density but no intrinsic curvature. 
\item Situations in which the loop is not spanned by a fluid film and the filament it is made from has no intrinsic curvature can be realized by setting both the surface tension $\sigma$ and the intrinsic curvature $\kappa_0$ to zero. By the third of these requirements, \eqref{etasolution}$_3$ holds trivially and \eqref{noten} reduces to
\be
\mu^2
=n^2 - 1,
\qquad
n\ge2.
\label{musolutionnoic}
\ee
%
For $n=2$, \eqref{musolutionnoic} coincides with \citepos{Michell} condition
\be
|\mu|=\sqrt{3}
\ee
for the buckling of a loop of length $2\pi R$ made from a filament with bending rigidity $a>0$, twisting rigidity $c>0$, and intrinsic twist density $\omega_0\ne0$. This result demonstrates that a loop made from a filament with intrinsic twist density but no intrinsic curvature will adopt a stable plane circular configuration unless the intrinsic twist density exceeds a certain threshold. 
%
%
\item Situations in which the loop is not spanned by a fluid film and the filament it is made from has no intrinsic twist density can be realized by setting the surface tension $\sigma$ and the intrinsic twist density $\omega_0$ equal to zero. By the first and second of these requirements, \eqref{etasolution}$_{1,2}$ coalesce to yield a condition
\be
k_n=1,
\qquad
n\ge2,
\label{musolutionnoitd}
\ee
equivalent to that derived by \cite{Goriely1} for the buckling of a loop made from a filament of length $2\pi R$ with bending rigidity $a>0$, twisting rigidity $c>0$, and intrinsic curvature $\kappa_0>0$. With reference to \eqref{ksubn}, for any $n\ge2$, $k_n$ increases monotonically as a function of $\beta$ for any choice of $\alpha>0$ but decreases monotonically with $\alpha$ for any choice of $\beta>0$. It therefore follows that increasing twist-to-bend ratio can counteract the destabilizing influence of intrinsic curvature.
\item Situations in which the loop is spanned by a fluid film but is made from the filament without intrinsic curvature can be realized by requiring that the surface tension $\sigma$ is positive and that the intrinsic curvature $\kappa_0$ vanishes. Since, by the second of these requirements, $\beta=R\kappa_0=0$, \eqref{ksubn} reduces to $k_n=0$, \eqref{etasolution}$_3$ holds trivially, and  \eqref{etasolution}$_{1,2}$ reduce to conditions
\be
\nu=\frac{n+1}{2} \left( 2n-1- \sqrt{1 + \frac{4\mu^2n}{n+1}}\,\right), 
\qquad 
\mu^2\le n^2 - 1,
\qquad
n\ge2,
\label{85}
\ee
equivalent to those derived by \cite{aisa2} for the stability of a loop made from a filament of length $2\pi R$ with bending rigidity $a>0$, twisting rigidity $c>0$, and intrinsic twist density $\omega_0$ spanned by a fluid film with surface tension $\sigma$. These results show that the presence of intrinsic twist density destabilizes an otherwise stable flat circular configuration, leading to buckling at values of the dimensionless surface tension $\eta=R^3\sigma/a$ below the critical value
\be
\nu=n^2-1,
\qquad
n\ge2,
\label{cfcond}
\ee
that arises for a loop made from a filament of length $2\pi R$ with bending rigidity $a>0$, $c=0$, $\kappa_0=0$, and $\omega_0=0$ derived by \cite{chen}.  
%
%
%
\end{itemize}


Returning to the general condition \eqref{etasolution}, for $n\ge2$ the corresponding buckling modes are determined by solving the systems \eqref{ad} and \eqref{bc} in conjunction with the linearized consequence \eqref{vconstraint} of the constraint \eqref{xiconstraint} of inextensibility. These solutions give the shape of the curve spanned by a surface with altitude $w$ determined (along with $v$, $\Psi$, and $\epsilon$) by solving the Laplace equation \eqref{diskshape} subject to the boundary conditions \eqref{rodshape1}--\eqref{rodshape2} and consequently deliver both a solution to the linearized version of the equilibrium conditions and the fundamental modes that bifurcate from the family of trivial solutions. Moreover, the projection of the surface onto the plane with unit normal $\bfe\times\bfe^{\sperp}$ is a unit disk only if $\bfv\cdot\bfe^{\sperp}=0$. Otherwise the projection is a planar region with noncircular boundary. If \eqref{etasolution} holds, the buckling solutions generally exhibit coupling between the in-plane and out-of-plane modes. If the twisting rigidity and intrinsic twist density of the filament that the loop is made from obey $c>0$ and $\omega_0\ne0$ (or, equivalently, by \eqref{alphaetc}$_1$ and \eqref{mu}, if $\alpha>0$ and $\mu\ne0$), the restriction $\bfv(1,\cdot)$ of $\bfv$ to the unit circle is given by
\begin{multline}
\bfv(1,\theta) = \left ( a_n(1) \cos  n\theta + b_n(1) \sin  n\theta \right ) \bfe + \frac{1}{n} \left ( b_n(1) \cos  n\theta - a_n(1) \sin  n\theta \right ) \bfe^{\sperp} \\
+  \frac{n+1}{2n \mu} \left ( 1 + k_n n - \sqrt{(1 - k_n n)^2 + \frac{4\mu^2n}{n+1}}\mskip4mu \right ) \left ( b_n(1) \cos  n\theta - a_n(1) \sin  n\theta \right )\bfe\times\bfe^{\sperp},
\label{bvsolution}
\end{multline}
where it is important to be cognizant that only one of the modal amplitudes $a_n(1)$ and $b_n(1)$ in \eqref{bvsolution} can be chosen independent of the other. 
%
%
If, alternatively, $c>0$ and $\omega_0=0$ (or, equivalently, by \eqref{alphaetc}$_1$ and \eqref{mu}, if $\alpha>0$ and $\mu=0$), then the stability condition \eqref{stabilitycriteria} and the buckling condition \eqref{etasolution} combine to yield
%
%
\be
\eta=\begin{cases}
n^2 - 1, & 0\le nk_n\le1, 
\\[4pt]
n(n+1)(1 - k_n), & 1\le nk_n\le n,
\end{cases}
\label{etaspecial}
\ee
in which case the in-plane and out-of-plane buckling modes decouple completely and \eqref{bvsolution} is replaced by
\be
\bfv(1,\theta)=
\begin{cases}
\displaystyle
( a_n(1) \cos  n\theta + b_n(1) \sin n \theta  ) \bfe + \frac{1}{n}  ( b_n(1) \cos  n\theta - a_n(1) \sin  n\theta  ) \bfe^{\sperp},
& 0 \le nk_n \le 1,
%
%
\\[6pt]
(c_n(1) \cos  n\theta + d_n(1) \sin  n\theta  ) \bfe \times \bfe^{\sperp},
& 1\le nk_n \le n.
\end{cases}
\label{inoutsolution}
\ee
%
%
%
%
Comparing \eqref{bvsolution} and \eqref{inoutsolution}, we see that the projection of \eqref{bvsolution} onto the plane with unit normal $\bfe\times\bfe^{\sperp}$ concurs with the purely in-plane expression \eqref{inoutsolution}$_1$ that arises if the filament that the loop is made from has no intrinsic twist density. Moreover, only in the case of the purely out-of-plane expression \eqref{inoutsolution}$_2$ do the projections of the buckling modes onto the plane with unit normal $\bfe\times\bfe^{\sperp}$ coincide with the unit disk. 
%
%

\subsection{Results for the most unstable buckling mode $n=2$}

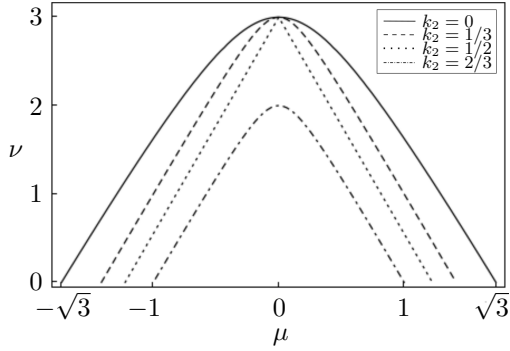
\begin{figure}
\begin{center}
\includegraphics[width = 66.67mm]{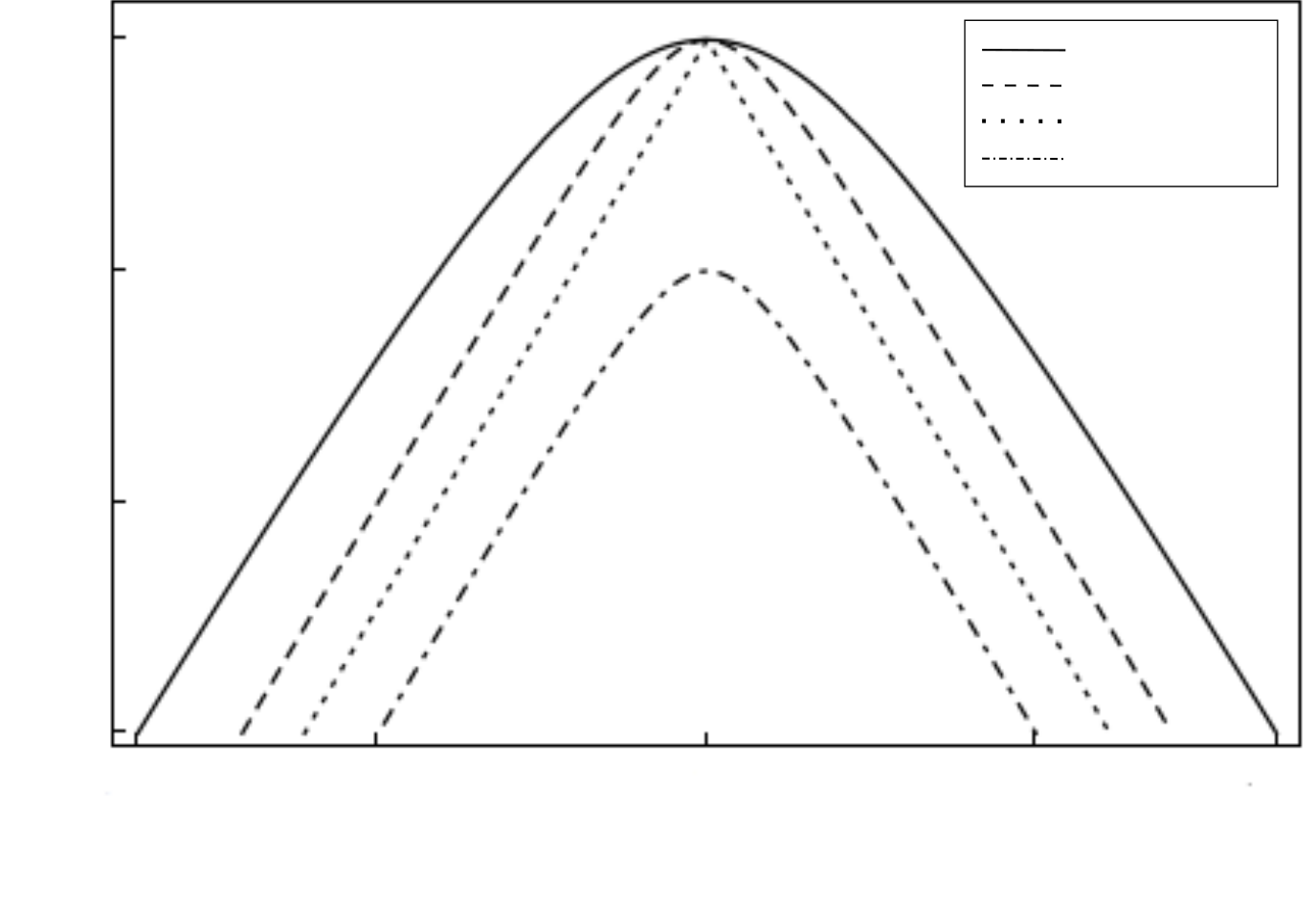}
\put(-33.5,121.5){\small \scalebox{0.75}{$k_2=0$}}
\put(-33.5,116.5){\small \scalebox{0.75}{$k_2=1/3$}}
\put(-33.5,111){\small \scalebox{0.75}{$k_2=1/2$}}
\put(-33.5,106){\small \scalebox{0.75}{$k_2=2/3$}}
\put(-180,12){$-\sqrt{3}$}
\put(-145,12){$-1$}
\put(-90,12){$0$}
\put(-43,12){$1$}
\put(-14,12){$\sqrt{3}$}
\put(-182, 22){$0$}
\put(-182,55){$1$}
\put(-182,88){$2$}
\put(-182,122){$3$}
\put(-90,3){$\mu$}
\put(-190,72){$\nu$}
\caption{Buckling diagram for collection of $(\mu,\eta)$ pairs for the most unstable buckling mode $n=2$ for different values of $k_2=(\alpha+\beta)\beta/(4\alpha+\beta)$.} 
\label{BucklingDiagram}
\end{center}
\end{figure}

The buckling condition \eqref{etasolution} must hold for all choices of the mode number $n\ge2$. For the most unstable buckling mode $n=2$, the contents of \eqref{etasolution} for various values of $k_2$ are provided in Figure~\ref{BucklingDiagram}.
The presence of intrinsic curvature and intrinsic twist density make elements of the trivial solution branch less stable, leading to buckling at values of $\eta$ below the critical value $\eta=3$ for the problem in which the rest configuration of the filament is rectilinear. 

\begin{figure}
\begin{center}
\includegraphics[width = 66.67mm]{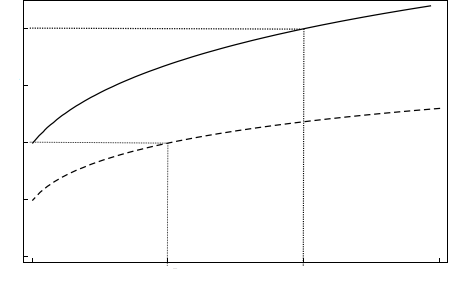}
\put(-187, 10){$0$}
\put(-195, 33){$0.5$}
\put(-187, 57){$1$}
\put(-195,81){$1.5$}
\put(-187,105){$2$}
\put(-205, 57){$\beta$}
\put(-178, 1){$0$}
\put(-126, 1){$0.5$}
\put(-64, 1){$1$}
\put(-12, 1){$1.5$}
\put(-95,-10){$\alpha$}
\put(-110,80){\scalebox{0.85}{$\Delta_{\text{out-of-plane}}$}}
\put(-110,35){\scalebox{0.85}{$\Delta_{\text{in-plane}}$}}
\caption{Illustration of the separation between in-plane and out-of-plane buckling regions in the $(\alpha,\beta)$-plane for a filament without intrinsic twist density and the most unstable buckling mode $n=2$.} 
\label{Decoupling_Buckling}
\end{center}
\end{figure}
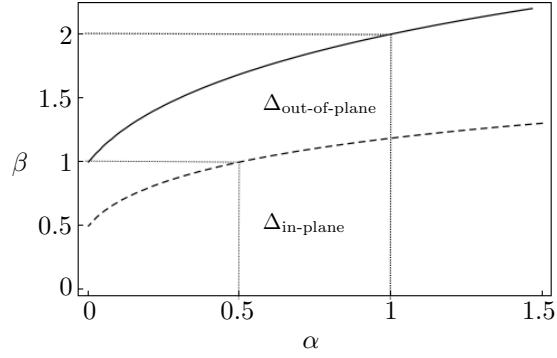

The effect described above is evident even in the case where the rest configuration of the filament has no intrinsic twist density, so that $\omega_0=0$ (or, equivalently, by \eqref{alphaetc}$_3$, $\gamma=0$). Under these circumstances, since, by \eqref{ksubn},
\be
k_2=\frac{(\alpha+\beta)\beta}{4\alpha+\beta}, 
\label{k2}
\ee
we deduce from \eqref{etaspecial} that:
\begin{itemize}
\item In-plane buckling occurs at a value $\nu=3$ of the dimensionless surface tension for all ordered combinations $(\alpha,\beta)$ of the twist-to-bend ratio and the dimensionless intrinsic curvature belonging to the region 
\be
\Delta_{\text{in-plane}}=\{(\alpha,\beta):\alpha>0,\beta\ge0,0\le2(\alpha+\beta)\beta<4\alpha+\beta\}.
\label{inplane}
\ee
\item Out-of-plane buckling occurs at the value
\be
\eta=6\bigg(1-\displaystyle\frac{(\alpha+\beta)\beta}{4\alpha+\beta}\bigg)<3
\label{etaspec}
\ee
of the dimensionless surface tension for all ordered combinations $(\alpha,\beta)$ of the twist-to-bend ratio and the dimensionless intrinsic curvature belonging to the region
\be
\Delta_{\text{out-of-plane}}=\{(\alpha,\beta):\alpha>0,4\alpha+\beta\le2(\alpha+\beta)\beta<2(4\alpha+\beta)\}.
\label{outofplane}
\ee
\end{itemize}
For all other values of $\alpha>0$ and $\beta\ge0$, the trivial solution is otherwise unstable if the loop is not spanned by a fluid film, namely if $\sigma=0$ (or, equivalently, by \eqref{alphaetc}$_4$, if $\eta=0$). These results are summarized in Figure~\ref{Decoupling_Buckling} and have the following consequences:
\begin{itemize}
\item If the ordered combination $(\alpha,\beta)$ of the twist-to-bend ratio and the dimensionless intrinsic curvature belongs to $\Delta_{\text{in-plane}}$, then the trivial solution branch undergoes a stable bifurcation to a flat noncircular solution branch at the value $\eta=3$ of the dimensionless surface tension. In view of \eqref{inplane}, this occurs only if 
\be
\beta\le\frac{1}{2}\bigg(\sqrt{\alpha^2+7\alpha+\frac{1}{4}}-\alpha+\frac{1}{2}\bigg).
\label{kappanaught}
\ee
For each positive choice of the twist-to-bend ratio $\alpha$, the critical value of the dimensionless surface tension $\eta$ at which a stable bifurcation from the trivial solution branch to the flat noncircular solution branch occurs is accordingly identical to that derived by \cite{chen} for the particular version of our problem in which the rest configuration of the filament from which the bounding loop is made is rectilinear. From \eqref{kappanaught}, the maximum value of the dimensionless intrinsic curvature $\beta$ allowed before bifurcation occurs increases monotonically with the twist-to-bend ratio $\alpha$. Thus, with reference to the definitions \eqref{alphaetc}$_{1,2}$, increasing the value of the twisting rigidity $c$ relative to the value of the bending rigidity $a$ increases the value of the intrinsic curvature $\kappa_0$ at which the bifurcation from the trivial solution branch to the flat noncircular solution branch occurs. 
\item If the ordered combination $(\alpha,\beta)$ of the twist-to-bend ratio and the dimensionless intrinsic curvature belongs to $\Delta_{\text{out-of-plane}}$, then, making reference to the specialization of \eqref{inoutsolution}$_2$ that arises for $n=2$, the trivial solution branch undergoes a stable bifurcation to a saddle-like out-of-plane solution branch at a critical value of the dimensionless surface tension $\eta$ determined by \eqref{etaspec} and thus below the critical value of associated with the stable bifurcation to the flat noncircular solution branch for $(\alpha,\beta)$ belonging to $\Delta_{\text{in-plane}}$. In view of \eqref{outofplane}, this occurs only if
\be
\frac{1}{2}\bigg(\sqrt{\alpha^2+7\alpha+\frac{1}{4}}-\alpha+\frac{1}{2}\bigg) \le \beta \le \frac{1}{2}\bigg(\sqrt{\alpha^2+14\alpha+1} - \alpha + 1\bigg).
\label{kappanaught1}
\ee
From \eqref{etaspec}, the critical value of the dimensionless surface tension $\eta$ increases monotonically with the twist-to-bend ratio $\alpha$ for any choice of the dimensionless intrinsic curvature satisfying $\beta>0$ but decreases monotonically with $\beta>0$ for any choice of $\alpha$, and from \eqref{kappanaught1}, the permissible range for the dimensionless intrinsic curvature (such as the difference between upper and lower bounds in \eqref{kappanaught1} for $\beta$) increases monotonically with $\alpha$. Hence, with reference to the definitions \eqref{alphaetc}$_{1,2}$, the destabilizing influence of the intrinsic curvature $\kappa_0$ can be countered by increasing the twisting rigidity $c$ relative to the bending rigidity $a$.

\item  If attention is restricted to the purely planar version of the problem, so that out-of-plane buckling cannot occur, then the trivial solution branch exhibits a stable bifurcation to a flat noncircular solution branch at the critical value $\eta=3$ of the dimensionless surface tension regardless of the values of the twist-to-bend ratio $\alpha>0$ and the dimensionless intrinsic curvature $\beta\ge0$. As a hydrostatically pressurized circular ring confined to two dimensions (or an infinitely long cylindrical pipe under pressure) is an equivalent example for the class of our planar problem; thus, we recover a result due to \cite{Katifori} which states that critical value of the dimensionless pressure difference in that context is not affected by the presence of intrinsic curvature.
\end{itemize}

\begin{figure}
\begin{center}
\includegraphics[width = 66.67mm]{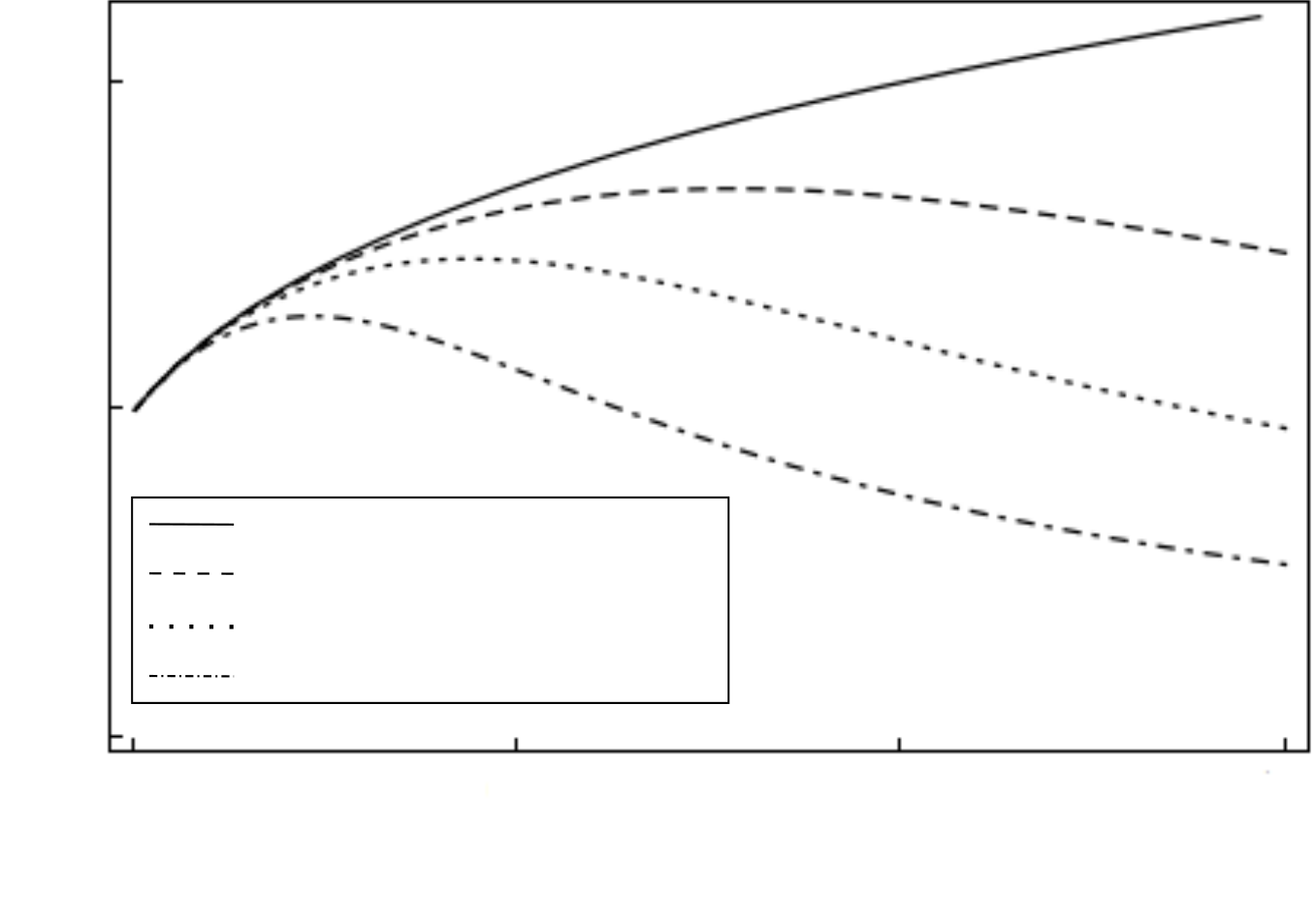}
\put(-182, 21){$0$}
\put(-182, 67){$1$}
\put(-182,115){$2$}
\put(-198, 67){$\beta$}
\put(-173, 11){$0$}
\put(-121, 11){$0.5$}
\put(-62, 11){$1$}
\put(-11, 11){$1.5$}
\put(-93,-2){$\alpha$}
\put(-153.0,52){\small \scalebox{0.75}{$\omega_0=0$}}
\put(-153.0,45.5){\small \scalebox{0.75}{$|\omega_0|=\kappa_0/2, \alpha_*=0.78$}}
\put(-153.0,37.75){\small \scalebox{0.75}{$|\omega_0|=\kappa_0, \alpha_*=0.44$}}
\put(-153.0,30){\small \scalebox{0.75}{$|\omega_0|=2\kappa_0, \alpha_*=0.23$}}
\caption{Illustration of coupling between in-plane and out-of-plane buckling regions in the $(\alpha,\beta)$-plane for a filament with various values of the ratio $|\omega_0|/\kappa_0$ with $\kappa_0>0$ and the most unstable buckling mode $n=2$. The value of $\alpha_*$ corresponding to the maximum value of $\beta$ for various values of the ratio $|\omega_0|/\kappa_0$ with $\omega_0\ne0$ is also indicated on the plot.} 
\label{Twist_Buckling}
\end{center}
\end{figure}
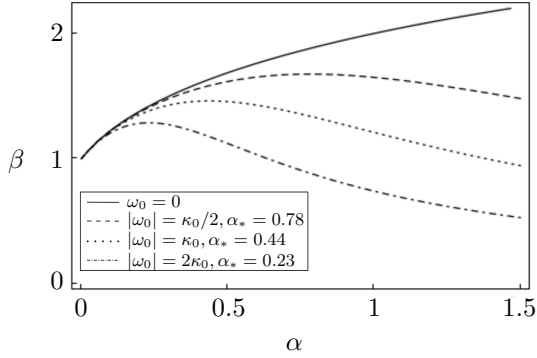

%
%
%
%
%
If the loop is made from a filament with positive intrinsic curvature $\kappa_0$ and nonvanishing intrinsic twist density $\omega_0$, then we deduce from \eqref{alphaetc}$_{1-3}$, \eqref{mu},  \eqref{ksubn}, and \eqref{etasolution} that, for all ordered combinations $(\alpha,\beta)$ of the twist-to-bend ratio and the dimensionless intrinsic curvature belonging to the region
\be
\Delta_\text{coupling}=\{(\alpha,\beta):\alpha>0,\beta\ge0,0\le3(\alpha+\beta)\beta<(4\alpha+\beta)(3-\alpha^2\beta^2\omega_0^2/\kappa_0^2)\}
\label{coupling}
\ee
buckling involving coupling between in-plane and out-of-plane modes occurs at the value  
\be
\eta= 3\left( \frac{3}{2} - \displaystyle\frac{(\alpha+\beta)\beta}{4\alpha+\beta} - \sqrt{{\displaystyle\Big(\frac{1}{2} - \frac{(\alpha+\beta)\beta}{4\alpha+\beta}\Big)^2} 
+ \frac{2\alpha^2\beta^2\omega_0^2}{3\kappa_0^2}}\,\right)
\label{etasolutionspec}
\ee
of the dimensionless surface tension. %
A diagram expressing the contents of \eqref{coupling} for various values of the ratio $|\omega_0|/\kappa_0$ with $\kappa_0>0$ is provided in Figure~\ref{Twist_Buckling}. In view of \eqref{coupling}, the curves in this diagram are determined by the relation
\be
\beta=\frac{1}{2}\bigg(\sqrt{\alpha^2+14\alpha\Big(1-\frac{\alpha^2\beta^2\omega_0^2}{3\kappa_0^2}\Big)+\Big(1-\frac{\alpha^2\beta^2\omega_0^2}{3\kappa_0^2}\Big)^2} - \alpha + 1 - \frac{\alpha^2\beta^2\omega_0^2}{3\kappa_0^2}\bigg).
\label{kappanaughtcoup}
\ee
%
%
%
It is evident from Figure~\ref{Twist_Buckling} that: 
\begin{itemize}
\item
The critical value of the dimensionless intrinsic curvature at which the flat circular configuration becomes unstable, with or without the liquid film, decreases monotonically with $|\omega_0|/\kappa_0$ for any choice of twist-to-bend ratio $\alpha>0$. It also follows from \eqref{etasolutionspec} that the critical value of the dimensionless surface tension $\eta$ decreases monotonically with $|\omega_0|/\kappa_0$ for any permissible value of the dimensionless intrinsic curvature $\beta\ge0$ and for any $\alpha>0$. Thus the presence of intrinsic twist density destabilizes the flat circular configuration, leading to buckling at values of the dimensionless surface tension $\eta$ below the critical value $\eta=3$ corresponding to the version of our problem in which the rest configuration of the filament is rectilinear. 

\item
If the filament is short enough to ensure that $\beta\le1$,
then the critical value of the dimensionless intrinsic curvature at which the circular shape of the filament becomes unstable with or without the liquid film decreases monotonically with the twist-to-bend ratio $\alpha>0$ for any choice of $|\omega_0|/\kappa_0$. Thus, in contrast to what occurs for a filament with no intrinsic twist density, where the twist-to-bend ratio $\alpha$ neutralizes the negative influence of the intrinsic curvature on the stability of the flat circular configuration, the destabilizing influence of the intrinsic curvature \textit{cannot} be countered by increasing $\alpha$. 
%
\item If the filament is long enough to ensure that $\beta > 1$,
then the critical value of the dimensionless intrinsic curvature at which the flat circular configuration becomes unstable increases monotonically with $\alpha$ satisfying $0<\alpha<\alpha_*$ and decreases monotonically with $\alpha$ satisfying $\alpha>\alpha_*$, for any choice of $|\omega_0|/\kappa_0$ and this is the case regardless of whether the loop is spanned by a fluid film.  The value $\alpha_*$ corresponding to the maximum value $\beta_*$ of $\beta$ is determined from the condition
\be
\frac{\text{d}\beta}{\text{d}\alpha} = 0
\label{Eb}
\ee
stipulating that the slope of the curve \eqref{kappanaughtcoup} vanishes.
It can be shown from \eqref{etasolutionspec}--\eqref{Eb} that for every permissible value of dimensionless intrinsic curvature $\beta$ satisfying $1<\beta<\beta_*$, there exists a value $\alpha_0$ of the twist-to-bend ratio ($0<\alpha_0<\alpha_*$) such as the critical value of $\eta$ at which a stable bifurcation 
occurs increases monotonically with $\alpha$ for $0<\alpha<\alpha_0$ and decreases monotonically with $\alpha$ for $\alpha_0<\alpha<\alpha_*$ and that this holds for all choices of $|\omega_0|/\kappa_0>0$. It thus follows that both the critical value of the dimensionless surface tension $\eta$ at which bifurcation occurs and the permissible value of the dimensionless intrinsic curvature increase monotonically with the twist-to-bend ratio $\alpha$ satisfying $0<\alpha<\alpha_0$, and that this is the case for any choice of the $|\omega_0|/\kappa_0>0$. 
We therefore arrive at the conclusion that, in contrast to what occurs for a filament with no intrinsic twist density, where the twist-to-bend ratio $\alpha$ always has the positive effect on the stability of the flat circular configuration, the twist-to-bend ratio $\alpha$ in this case has limited stabilizing effect.
\end{itemize}



\section{Summary and discussion}
\label{summary}

We have used a variational model to study the stability and buckling behavior of a loop spanned by a liquid film. In contrast to prior works, we considered loops made from filaments with intrinsic curvature and twist. Consistent with prior works, however, we restricted attention to filaments with  circular cross sections of constant diameter and uniform material properties. Coupling between the bending of its midline and the orientation of its cross sections is incorporated in a simple way that, among other things, distinguishes between bending toward and away from the center of intrinsic curvature. Our model involves five material parameters: the bending rigidity $a>0$, twisting rigidity $c>0$, intrinsic curvature $\kappa_0\ge0$, and intrinsic twist density $\omega_0$ of the filament along with the surface tension $\sigma$ of the liquid film. In combination with the length $2\pi R$ of the loop, these quantities give rise to four dimensionless parameters: the twist-to-bend ratio $\alpha=c/a$, the dimensionless intrinsic curvature $\beta=R\kappa_0$, the dimensionless intrinsic twist density $\gamma=R\omega_0$, and the dimensionless surface tension $\nu=R^3\sigma/a$. However, dependence on the dimensionless intrinsic twist density $\gamma$ occurs only through the product $\mu=\alpha\gamma$. Since the most easily tuned parameter in the theory is the length $2\pi R$ of the bounding loop, we hereafter express our results in terms of the dimensional parameter $R$. Moreover, we restrict discussion to the most unstable mode, namely $n=2$.

The effect of coupling between the bending of its midline and the orientation of its cross sections via the intrinsic curvature is evident for the particular case in which the bounding loop is made from a filament with intrinsic curvature but no intrinsic twist density. In this context, we find that there exist in-plane and out-of-plane buckling modes corresponding to stable solution branches that bifurcate from the branch of flat circular solutions. Specifically, with reference to \eqref{alphaetc}$_{1,2}$ and \eqref{kappanaught}, a flat circular configuration involving a loop  made from a filament with bending rigidity $a>0$, twisting rigidity $c>0$, and intrinsic curvature $\kappa_0>0$ undergoes a stable bifurcation to a flat noncircular configuration at a critical value of $\eta=R^3\sigma/a=3$ of the dimensionless surface tension if its radius obeys 
\be
R\le
\frac{1}{2\kappa_0}\bigg(\sqrt{\alpha^2+7\alpha+\frac{1}{4}}-\alpha+\frac{1}{2}\bigg),
\qquad
\alpha=\frac{c}{a}.
\label{kappanaughtbis}
\ee
Alternatively, with reference to \eqref{alphaetc}$_{1.2}$ and \eqref{kappanaught1}, a flat circular configuration involving a loop made from a filament with bending rigidity $a>0$, twisting rigidity $c>0$, and intrinsic curvature $\kappa_0>0$ undergoes a stable bifurcation to an out-of-plane configuration at a critical value of $\eta=R^3\sigma/a<3$ if its radius obeys
\be
\frac{1}{2\kappa_0}\bigg(\sqrt{\alpha^2+7\alpha+\frac{1}{4}}-\alpha+\frac{1}{2}\bigg)
\le 
R
\le
\frac{1}{2\kappa_0}\bigg(\sqrt{\alpha^2+14\alpha+1} - \alpha + 1\bigg),
\qquad
\alpha=\frac{c}{a}.
\label{kappanaught1bis}
\ee
Moreover, consistent with \eqref{kappanaughtbis} and \eqref{kappanaught1bis}, the destabilizing influence of the intrinsic curvature $\kappa_0$ can be countered by increasing the torsional rigidity $c$ relative to the bending rigidity $a$. 
%

\noindent Bearing in mind that for $\alpha>0$, 
\be
1\le\sqrt{\alpha^2+7\alpha+\frac{1}{4}}-\alpha+\frac{1}{2}<4, 
\label{LU}
\ee
it follows from \eqref{kappanaughtbis} that
a flat circular configuration involving a loop made from a filament of length $2\pi R$ with the twist-to-bend ratio $\alpha$ and intrinsic curvature $\kappa_0$ will buckle to a stable, flat noncircular configuration at dimensionless surface tension $\eta=R^3\sigma/a=3$, if the filament is short enough so that
\be
R\le\frac{1}{2\kappa_0}.
\label{loR}
\ee
%
Importantly, the foregoing conclusion holds independent of the twist-to-bend ratio $\alpha$ of the filament. When \eqref{loR} is violated, it is necessary to check whether \eqref{kappanaughtbis} or \eqref{kappanaught1bis} is satisfied for a specific value of $\alpha$ to determine whether in-plane or out-of-plane buckling arises. 
With reference to the Figure~\ref{Decoupling_Buckling}, however, we can conclude that in-plane or out-of-plane buckling must arise for certain ranges of values of the twist-to-bend ratio and the radius of the loop. Specifically, with reference to \eqref{LU} and observing that the respective values of the upper-bound of the right-hand side of \eqref{kappanaughtbis} for $\alpha=0.5$ and $\alpha=1$ are $\displaystyle1/\kappa_0$ and $\displaystyle(\sqrt{33}-1)/(4\kappa_0)$ respectively, it follows that:
%
\begin{itemize}
\item A flat circular configuration involving a loop made from a filament of length $2\pi R$ with bending rigidity $a>0$, twisting rigidity $c>0$, and intrinsic curvature $\kappa_0>0$ will buckle to a stable out-of-plane configuration at dimensionless surface tension $\eta=R^3\sigma/a<3$ if
\be
R \ge \frac{1}{\kappa_0} \qquad \text{and} \qquad0<c\le\frac{a}{2}.
\label{Op1}
\ee
\item A flat circular configuration involving a loop made from a filament of length $2\pi R$ with bending rigidity $a>0$, twisting rigidity $c>0$, and intrinsic curvature $\kappa_0$ will buckle either to a stable flat noncircular configuration at dimensionless surface tension $\eta=R^3\sigma/a=3$ if
\be
R \le \frac{1}{\kappa_0} \qquad \text{and} \qquad \frac{a}{2}<c\le a,
\ee
or to a stable out-of-plane configuration at dimensionless surface tension $\eta=R^3\sigma/a<3$ if
\be
R \ge \frac{\sqrt{33}-1}{4\kappa_0} \qquad \text{and} \qquad \frac{a}{2}<c\le a.
\ee
%
%
\item A flat circular configuration involving a loop made from a filament of length $2\pi R$ with bending rigidity $a>0$, twisting rigidity $c>0$, and intrinsic curvature $\kappa_0$ will buckle either to a stable flat noncircular configuration at dimensionless surface tension $\eta=R^3\sigma/a=3$ if
\be
R \le \frac{\sqrt{33}-1}{4\kappa_0} \qquad \text{and}  \qquad a<c,
\ee
or to a stable out-of-plane configuration at dimensionless surface tension $\eta=R^3\sigma/a<3$ if
\be
R \ge \frac{2}{\kappa_0} \qquad \text{and} \qquad a<c.
\label{Op2}
\ee
\end{itemize}
%

Suppose, now, that the filament is isotropic and linearly elastic with Young's modulus $E$ and shear modulus $G$. Recall that these moduli are related through the Poisson ration $\xi$ by $E=2(1+\xi)G$. Thus, since the cross section of the filament is assumed to be circular, the polar moment of inertia $J$ of the cross-section is twice the moment of inertia $I$ of the cross section: $J=2I$. Under these circumstances, the twist-to-bend ratio $\alpha=c/a$ has the form
\be
\alpha=\frac{c}{a}=\frac{GJ}{EI}=\frac{1}{1+\xi},
\qquad
-1<\xi\le\frac{1}{2},
\label{alpharestriction}
\ee
and the conditions \eqref{kappanaughtbis} and \eqref{kappanaught1bis} that are sufficient to induce stable in-plane and out-of-plane bifurcations from the flat circular configurations specialize to
\be
R\le\frac{\sqrt{\xi^2+30\xi+33}+\xi-1}{4\kappa_0(1+\xi)}
\label{suff1}
\ee
and
\be
\frac{\sqrt{\xi^2+30\xi+33}+\xi-1}{4\kappa_0(1+\xi)}
\le
R<\frac{\sqrt{\xi^2+16\xi+16}+\xi}{2\kappa_0(1+\xi)},
\label{suff2}
\ee
respectively. 
%
%
Determining the least upper-bound of the right-hand side of \eqref{suff1}, we deduce that a flat circular configuration involving a loop made from an isotropic linearly elastic filament of length $2\pi R$ with Poisson's ratio $\xi$ and intrinsic curvature $\kappa_0$ will buckle to a stable flat noncircular configuration at dimensionless surface tension $\eta=R^3\sigma/a=3$ if the filament is short enough to ensure that
\be
R\le\frac{\sqrt{193}-1}{12\kappa_0}.
\label{loR1}
\ee
Importantly, the foregoing conclusion holds independent of the Poisson's ratio $\xi$ of the filament. 
Employing reasoning analogous to that used to obtain results \eqref{Op1}--\eqref{Op2} and keeping in mind from \eqref{alpharestriction} that 
\be
\left.
\begin{aligned}
\displaystyle
\frac{a}{2} < c &\le a, \qquad \phantom{-}0\le\xi \le \frac{1}{2}, 
\\[4pt]
a &< c, \qquad -1<\xi < 0,
\end{aligned}
\,\right\}
\ee
it follows that:
\begin{itemize}
\item A flat circular configuration involving a loop made from a filament of length $2\pi R$ with the Poisson's ratio $\xi$ and intrinsic curvature $\kappa_0$ will buckle to a stable out-of-plane configuration at dimensionless surface tension $\eta=R^3\sigma/a<3$ if
\be
R \ge \frac{\sqrt{33}-1}{4\kappa_0} \qquad \text{and} \qquad 0\le\xi \le \frac{1}{2}.
\label{suff3}
\ee
%
%
\item 
A flat circular configuration involving a loop made from a filament of auxetic materials (with the Poisson's ratio $-1<\xi<0$) of length $2\pi R$  and intrinsic curvature $\kappa_0$ will buckle either to a stable, flat noncircular configuration at dimensionless surface tension $\eta=R^3\sigma/a=3$ if
\be
R \le \frac{\sqrt{33}-1}{4\kappa_0}, 
\ee
or to a stable out-of-plane configuration at dimensionless surface tension $\eta=R^3\sigma/a<3$ if
\be
R \ge \frac{2}{\kappa_0}. 
\ee
\end{itemize}
\noindent As an illustration, consider an isotropic linearly elastic filament of circular cross section, made from polyvinyl chloride (PVC) with an intrinsic radius of curvature $R_0=\kappa_0^{-1}$ and length $L=3\pi R_0$. Then $R=L/2\pi=3/2\kappa_0$ 
and since $\xi\approx0.31$ for PVC \citep{mohan}, 
quick calculations show that the condition \eqref{suff3} for a filament of the chosen dimensions and properties to buckle out-of-plane is satisfied. By \eqref{alpharestriction}, $\alpha=0.76$ and with reference to \eqref{etaspec}, the system therefore undergoes a stable bifurcation to a saddle-like, out-of-plane solution branch at the critical value of $\eta=R^3\sigma/a=3/2<3$ of the dimensionless surface tension. 

If, more generally, the bounding loop is made from a filament with both intrinsic curvature and intrinsic twist density, then, making reference to \eqref{alphaetc}$_{1,2}$, \eqref{mu}, \eqref{coupling}, and \eqref{kappanaughtcoup}, a flat circular configuration involving a loop made from a filament with bending rigidity $a>0$, twisting rigidity $c>0$, intrinsic curvature $\kappa_0>0$, and intrinsic twist density $\omega_0\ne0$ undergoes a stable bifurcation to configuration involving both in-plane and out-of-plane modes at a critical value of $\eta=R^3\sigma/a<3$ if its radius $R$ obeys 
\be
R\le\frac{\beta_c}{\kappa_0}, 
\label{kappanaughtcoup1}
\ee
where $\beta_c$ is the only positive solution of the equation \eqref{kappanaughtcoup}. Moreover, increasing the torsional rigidity $c$ relative to the bending rigidity $a$ has no or limited stabilizing effect if the loop is too short or too long.

Suppose, now, that the filament is isotropic and linearly elastic with Young's modulus $E$ and shear modulus $G$. As discussed previously, for such a material, the twist-to-bend ratio $\alpha=c/a$, from \eqref{alpharestriction}, satisfies
\be
\alpha\ge\frac{2}{3},
\qquad \text{for}\qquad
-1<\xi\le\frac{1}{2}.
\label{alphabound}
\ee
We now find the critical value of $|\omega_0|\kappa_0^{-1}$ above which increasing the torsional rigidity $c$ relative to the bending rigidity $a$ has no stabilizing effect on stability of flat circular configurations of loops made from filaments with intrinsic curvature $\kappa_0$ and intrinsic twist density $\omega_0$. As discussed previously, if the loop is sufficiently long to ensure that $R>\kappa_0^{-1}$, then the permissible value $\beta_c$ of the dimensionless intrinsic curvature decreases monotonically with the twist-to-bend ratio $\alpha$ satisfying $\alpha>\alpha_*$, with $\alpha_*$ being the solution of \eqref{Eb}. With reference to \eqref{alphabound}, it is thus sufficient to find the critical value of $|\omega_0|\kappa_0^{-1}$ from the condition that equation \eqref{Eb} has solution $\alpha_*=2/3$. Straightforward calculations yield $|\omega_0|\kappa_0^{-1}=0.61$. Since the value $\alpha_*$ corresponding to the maximum value of $\beta_c$ decreases monotonically with $\omega_0|\kappa_0^{-1}$, it  follows from \eqref{alphabound} that the permissible value $\beta_c$ of the dimensionless intrinsic curvature indeed decreases monotonically with the twist-to-bend ratio for filaments made from an isotropic linearly elastic material if $|\omega_0|\kappa_0^{-1}>0.61$ and the filaments are sufficiently long. Combining with the understanding that the admissible value $\beta_c$ of the dimensionless intrinsic curvature decreases monotonically with the twist-to-bend ratio for filaments that are short enough to ensure that $R<\kappa_0^{-1}$, and this is the case for all choices of the value of $|\omega_0|\kappa_0^{-1}$; thus, we arrive at the conclusion that increasing the torsional rigidity $c$ relative to the bending rigidity $a$ has no stabilizing effect on stability of flat circular configurations of loops made from an isotropic linearly elastic filament with intrinsic curvature $\kappa_0$ and intrinsic twist density $\omega_0$ if $|\omega_0|>0.61\kappa_0$, regardless of the length of the loop. It is evident that the above conclusion holds for filaments that possess intrinsic twist density but do not possess intrinsic curvature.


As an illustration, consider an isotropic, linearly elastic filament of circular cross section, made from polyvinyl chloride (PVC) with intrinsic radius of curvature $R_0=\kappa_0^{-1}$, length $L=3\pi R_0$, and intrinsic twist density $\omega_0{=\kappa_0/3}$ (in which case the filament has total twist $\int_0^L \omega_0\,\text{d}s = \pi$). Then, as in the previously discussed example, $\xi=0.31$ and $\alpha=0.76$. It therefore follows from \eqref{kappanaughtcoup} that \eqref{kappanaughtcoup1} is satisfied with $\beta_c=1.77$ for a filament of the chosen dimensions and properties. Evaluating the right-hand side of \eqref{etasolutionspec} accordingly, we infer that the system undergoes a stable bifurcation to a branch of solutions in which both in-plane and out-of-plane modes are represented at the critical value of $\eta=R^3\sigma/a=1.1<3$ of the dimensionless surface tension.

Being based on a linear analysis, our study provides no information on the shapes of post-buckled configurations. Additionally, our model is limited by the assumption that the bounding loop is free of self-contact. The importance of allowing for self-contact is evident from the literature on twisted elastic rings, where it is known that increasing the twist density of a circular ring above the threshold needed to induce buckling out-of-plane causes the ring to loop back on itself and form a twisted figure-eight with a point of self-contact. For future work, it therefore seems important to extend the model to account for the unilateral constraints needed to model self-contact and to perform detailed post-buckling studies.

\section*{Acknowledgment}

The authors gratefully acknowledge support from the Okinawa Institute of Science and Technology Graduate University with subsidy funding from the Cabinet Office, Government of Japan. They also thank Giulio Giusteri for insightful discussions and Steven Aird for helpful guidance on matters of grammar and style.

\section*{References}

\bibliography{Reference.bib}

\end{document}